\documentclass[aps,pra,amsmath,amssymb,reprint,groupedaddress,showpacs]{revtex4-1}

\usepackage{graphicx}
\usepackage{dcolumn}
\usepackage{bm}

\begin{document}


\title{Dirac electron in a chiral space-time crystal created by counterpropagating circularly polarized plane electromagnetic waves}


\author{G. N. Borzdov}
\email[]{BorzdovG@bsu.by}
\affiliation{Department of Theoretical Physics and Astrophysics, Belarusian State University,
4 Nezavisimosti Avenue, 220030 Minsk, Belarus}



\begin{abstract}
The family of solutions to the Dirac equation for an electron moving in an electromagnetic lattice with the chiral structure created by counterpropagating circularly polarized plane electromagnetic waves is obtained. At any nonzero quasimomentum, the dispersion equation has two solutions which specify bispinor wave functions describing electron states with different energies and mean values of momentum and spin operators. The inversion of the quasimomentum results in two other  linearly independent solutions. These four basic wave functions are uniquely defined by eight complex scalar functions (structural functions), which serve as convenient building blocks of the relations describing the electron properties. These properties are illustrated in graphical form over a wide range of quasimomenta. The superpositions of two basic wave functions describing different spin states and corresponding to (i) the same quasimomentum (unidirectional electron states with the spin precession) and (ii) the two equal-in-magnitude but oppositely directed quasimomenta (bidirectional electron states) are also treated.

\end{abstract}


\maketitle

\section{Introduction}
The motion of electrons in natural crystals is described by the Schr\"{o}dinger equation with a periodic electrostatic scalar potential. Electromagnetic fields with periodic dependence on space-time coordinates can be treated by analogy with the crystals of solid-state physics, so it is natural to refer to these field lattices as electromagnetic space-time crystals (ESTCs)~\cite{gaps,bian04,ESTCp1,ESTCp2,ESTCp3,ESTCp4}. In this context, the idea of a space-time crystal was first presented in \cite{gaps} and the electron wave functions for the ESTC, created by two linearly polarized plane waves, were calculated by using the first-order perturbation theory for the Schr\"{o}dinger-Stueckelberg equation. The terms ``time crystal" and ``space-time crystal" have been used previously in other contexts, in particular, in the recent discussion around the question of whether time-translation symmetry might be spontaneously broken in a time-independent, conservative classical system~\cite{cltime} and a closed quantum mechanical system~\cite{qtime}, such as ions confined in a ring-shaped trapping potential with a static magnetic field~\cite{ions,watanabe} or a one-dimensional chain of ytterbium ions~\cite{Yao}.

An electron in an electromagnetic field with the four-dimensional potential $\bm{A}=(\textbf{A},i \varphi)$ is described by the Dirac equation
\begin{equation}\label{Dirac}
   \left[\gamma_k\left(\frac{\partial}{\partial x_k} - i A_k\frac{e}{c \hbar}\right) + \kappa_{e}\right]\Psi=0,
\end{equation}
where $\kappa_{e}=m_{e} c/ \hbar$, $c$ is the speed of light in vacuum, $\hbar$ is the Planck constant, $e$ is the electron charge, $m_{e}$ is the electron rest mass,  $\gamma_k$ are the Dirac matrices, $\Psi$ is the bispinor, $x_1$, $x_2$, and $x_3$  are the Cartesian coordinates, $x_4=i c t$, and summation over repeated indices is carried out from 1 to 4.

In~\cite{ESTCp1,ESTCp2,ESTCp3,ESTCp4}, we obtained the fundamental solution of Eq.~(\ref{Dirac}) and presented tools for its numerical analysis in the case when $A_4\equiv i \varphi=0$ and
\begin{equation}\label{field}
  \textbf{A}^\prime \equiv \frac{e}{m_e c^2}\textbf{A}=\sum_{j=1}^6\left(\textbf{A}_j e^{i \bm{K}_j \cdot \bm{x}}+\textbf{A}_j^\ast e^{-i \bm{K}_j \cdot \bm{x}}\right).
\end{equation}
This ESTC is created by six plane waves with unit wave normals $\pm\textbf{e}_\alpha$, where $\textbf{e}_\alpha$ are the orthonormal basis vectors,  $\bm{x}=(\textbf{r},ict)$, $\textbf{r}=x_1\textbf{e}_1+x_2\textbf{e}_2+x_3\textbf{e}_3$. All six waves
have the same frequency $\omega_0$ and
\begin{equation}\label{KNj}
    \bm{K}_\alpha =(k_0 \textbf{e}_\alpha,i k_0), \quad \bm{K}_{\alpha +3}=(-k_0 \textbf{e}_\alpha ,i k_0),
\end{equation}
where $\alpha =1,2,3$,  and $k_0=\omega_0/c=2\pi/\lambda_0$. They may have any polarization, so that their complex amplitudes are specified by dimensionless real constants  $a_{jk}$ and $b_{jk}$ as follows:
\begin{equation}
    \textbf{A}_j = \sum_{k=1}^3 \left(a_{jk} + i b_{jk}\right)\textbf{e}_k, \quad j = 1,2,...,6,\label{abjk}
\end{equation}
where $a_{jj} = b_{jj} = a_{j+3\,j} = b_{j+3\,j} = 0, j=1,2,3$.

In the general case, Eqs.~(\ref{field})--(\ref{abjk}) describe a four-dimensional ESTC (4D-ESTC), i.e., with periodic dependence on all four space-time coordinates. The condition $\textbf{A}_3=\textbf{A}_6=0$ reduces it to a 3D-ESTC with periodic dependence on $x_1, x_2, x_4$, whereas the condition $\textbf{A}_2=\textbf{A}_3=\textbf{A}_5=\textbf{A}_6=0$ results in a 2D-ESTC periodic in $x_1, x_4$. In the simplest case, when $\textbf{A}_1$ is the only nonzero amplitude, Eq.~(\ref{Dirac}) has the well-known Volkov solution~\cite{Volkov}. There exist different representations of this solution~\cite{ESTCp4,Fed,Tern}.

The new technique presented in~\cite{bian04,ESTCp1,ESTCp2,ESTCp3,ESTCp4} is applied in~\cite{ESTCp3} to the 4D-ESTCs created by the linearly polarized waves with the amplitudes
\begin{eqnarray}
  \textbf{A}_1&=&-\textbf{A}_4=A_m\textbf{e}_2,\nonumber\\
  \textbf{A}_2&=&-\textbf{A}_5=A_m\textbf{e}_3,\nonumber\\
  \textbf{A}_3&=&-\textbf{A}_6=A_m\textbf{e}_1, \label{A16cr1}
\end{eqnarray}
and the circularly polarized waves with the amplitudes
\begin{eqnarray}\label{Acr2}
    \textbf{A}_1&=&\textbf{A}_4=A_m(\textbf{e}_2+i \textbf{e}_3)/\sqrt{2},\nonumber\\
    \textbf{A}_2&=&\textbf{A}_5=A_m(\textbf{e}_3+i \textbf{e}_1)/\sqrt{2},\nonumber\\
    \textbf{A}_3&=&\textbf{A}_6=A_m(\textbf{e}_1+i \textbf{e}_2)/\sqrt{2},
\end{eqnarray}
respectively, where $A_m$ is a real scalar amplitude. It is shown that the second one possesses the spin birefringence. In~\cite{ESTCp4}, this technique is illustrated by the analysis of the ground state and the spin precession of the Dirac electron in the field of two counterpropagating plane waves with left and right circular polarizations, i.e., in the 2D-ESTC with the nonzero amplitudes
\begin{equation}\label{A1A4}
    \textbf{A}_1=\textbf{A}_4=A_m(\textbf{e}_2+i \textbf{e}_3)/\sqrt{2}.
\end{equation}

In the present paper, we treat the electron motion in the chiral 2D-ESTC defined by the amplitudes $\textbf{A}_1=\textbf{A}_4^{\ast}=A_m(\textbf{e}_2+i \textbf{e}_3)$, so that
\begin{equation}\label{Aprime}
  \textbf{A}^\prime = 4 A_m \cos \varphi_4 \textbf{e}_A(\varphi_1),
\end{equation}
where
\begin{equation}\label{eA}
    \textbf{e}_A(\varphi_1)=\textbf{e}_2\cos\varphi_1-\textbf{e}_3\sin\varphi_1,
\end{equation}
and $\varphi_j=2\pi X_j, j=1,2,3,4; X_k=x_k/\lambda_0, k=1,2,3, X_4=ct/\lambda_0$. The interplay between the fundamental solution of Eq.~(\ref{Dirac}) and particular solutions, specified by given initial amplitudes, for the general 4D-ESTC and the chiral ESTC is discussed in Sec.~\ref{sec:baseq}. The four basic solutions which describe two different spin states of the Dirac electron moving in the 2D-ESTC along the $X_1$ axis in the positive and negative directions are presented in Sec.~\ref{sec:wave}. In Sec.~\ref{sec:family}, we treat superpositions of two basic wave functions describing different spin states and corresponding to (i) the same quasimomentum  (unidirectional electron states) and (ii) the two equal-in-magnitude but oppositely directed quasimomenta (bidirectional electron states). In the general 4D-ESTC, the Dirac equation reduces to an infinite system of matrix equations, where the interconnections between equations are defined~\cite{ESTCp3,ESTCp4} by 12 matrix functions and 56 scalar coefficients. The Appendix gives the expressions for them in an explicit form. In the chiral 2D-ESTC, the number of these interconnections decreases drastically, resulting in specific interrelations between the basic solutions discussed in Sec.~\ref{sec:structure}.

\section{\label{sec:baseq}Basic relations}
\subsection{Fundamental solution}
The electron wave function in the 4D-ESTC can be written as follows~\cite{ESTCp1,ESTCp4}:
\begin{equation}\label{sol1}
    \Psi=\Psi_0 e^{i \bm{x}\cdot\bm{K}},\quad \Psi_0=\sum_{n \in \mathcal L} c(n) e^{i \bm{x}\cdot\bm{G}(n)},
\end{equation}
where $\bm{K} = (\textbf{k},i \omega /c)$ is the four-dimensional wave vector, $\textbf{k} = k_1\textbf{e}_1+k_2\textbf{e}_2+k_3\textbf{e}_3$, $\bm{G}(n) = (k_0 \textbf{n},i k_0 n_4), \textbf{n} = n_1\textbf{e}_1+n_2\textbf{e}_2+n_3\textbf{e}_3$, points $n=(n_1,n_2,n_3,n_4)$ of the integer lattice $\mathcal L$ have even values of the sum $n_1+n_2+n_3+n_4$, and
\begin{equation}\label{cn}
    c(n) = \left(
             \begin{array}{c}
               c^1(n) \\
               c^2(n) \\
               c^3(n) \\
               c^4(n) \\
             \end{array}
           \right) \equiv \left(
           \begin{array}{c}
               c^1 \\
               c^2 \\
               c^3 \\
               c^4 \\
             \end{array}
           \right)_n
\end{equation}
are the Fourier amplitudes (bispinors). The function $\Psi_0$ is periodic in $X_1, X_2, X_3$, and $X_4$ with the unit period. At a given $\bm{K}$, the set of functions  $\Psi$~(\ref{sol1}) is the Hilbert space with the scalar product
\begin{eqnarray}\label{scalarproduct}
  (\Psi_a,\Psi_b)=&&\int_0^1dX_1\int_0^1dX_2\int_0^1dX_3\int_0^1dX_4 \Psi_a^{\dag}\Psi_b\nonumber\\
     =&&\sum_{n \in \mathcal L}a^{\dag}(n)b(n)
\end{eqnarray}
and the norm
\begin{equation}\label{normpsi}
    \|\Psi\|=(\Psi,\Psi)^{1/2}=\left(\sum_{n \in \mathcal L}c^{\dag}(n)c(n)\right)^{1/2},
\end{equation}
where
\begin{eqnarray}\label{psiapsib}
  \Psi_a&=&\Psi_{0a} e^{i \bm{x}\cdot\bm{K}},\quad \Psi_{0a}=\sum_{n \in \mathcal L} a(n) e^{i \bm{x}\cdot\bm{G}(n)},\\
  \Psi_b&=&\Psi_{0b} e^{i \bm{x}\cdot\bm{K}},\quad \Psi_{0b}=\sum_{n \in \mathcal L} b(n) e^{i \bm{x}\cdot\bm{G}(n)}.
\end{eqnarray}

Let us treat the infinite set $C =\{c(n),n \in {\mathcal L}\}$ of the Fourier amplitudes $c(n)$ of the wave function $\Psi$ (\ref{sol1}) as an element of an infinite-dimensional complex linear space $V_C$. Since for any given $n \in {\mathcal L}$, $c(n)$ is the bispinor, $C \in V_C$ will be called the multispinor. The basis $e_j(n)$ in $V_C$ and the dual basis $\theta^j(n) = e_j^{\dag}(n)$ in the space of one-forms $V_C^\ast$ are specified as follows:
\begin{eqnarray}\label{e14n}
  e_1(n)&=&\left(
             \begin{array}{c}
               1 \\
               0 \\
               0 \\
               0 \\
             \end{array}
           \right)_n, \quad e_2(n) = \left(
             \begin{array}{c}
               0 \\
               1 \\
               0 \\
               0 \\
             \end{array}
           \right)_n, \nonumber\\
  e_3(n)&=&\left(
             \begin{array}{c}
               0 \\
               0 \\
               1 \\
               0 \\
             \end{array}
           \right)_n, \quad e_4(n) = \left(
             \begin{array}{c}
               0 \\
               0 \\
               0 \\
               1 \\
             \end{array}
           \right)_n,
\end{eqnarray}
\begin{eqnarray}\label{t14n}
  \theta^1(n)&=&\left(
                          \begin{array}{cccc}
                            1 & 0 & 0 & 0 \\
                          \end{array}
                        \right)_n, \quad \theta^2(n) = \left(
                          \begin{array}{cccc}
                            0 & 1 & 0 & 0 \\
                          \end{array}
                        \right)_n,\nonumber  \\
  \theta^3(n)&=&\left(
                          \begin{array}{cccc}
                            0 & 0 & 1 & 0 \\
                          \end{array}
                        \right)_n, \quad \theta^4(n) = \left(
                          \begin{array}{cccc}
                            0 & 0 & 0 & 1 \\
                          \end{array}
                        \right)_n,
\end{eqnarray}
that is,
\begin{eqnarray}\label{timejn}
  \left\langle \theta^i(m),e_j(n)\right\rangle&=&1 \text{ for } m=n \text{ and } i=j \nonumber\\
                                           &=&0 \text{ for } m\neq n \text{ and/or } i\neq j
\end{eqnarray}
for any $m, n \in {\mathcal L}$ and $i,j=1,2,3,4$. The unit operator $\mathcal{U}$ in $V_C$ can be written as
\begin{equation}\label{UVC}
    \mathcal{U}=\sum_{n\in \mathcal{L}} I(n),\quad I(n)=e_j(n)\otimes\theta^j(n), \quad tr[I(n)]=4.
\end{equation}

Substitution of $\textbf{A}$ (\ref{field}) and $\Psi$ (\ref{sol1}) in Eq.~(\ref{Dirac}) results in the infinite system of matrix equations~\cite{ESTCp1,ESTCp4},
\begin{equation}\label{meq}
\sum_{s \in S_{13}} V(n,s)c(n+s)=0, \quad {n\in \mathcal{L}},
\end{equation}
where $s=(s_1,s_2,s_3,s_4)$ satisfies the condition $g_{4d}(s)=0, 1$, $g_{4d}(s_1,s_2,s_3,s_4)=\max\{|s_1|+|s_2|+|s_3|,|s_4|\}$, i.e.,
\begin{eqnarray}\label{S13}
s\in S_{13}=&&\{s_h(i),i=0,1,...,12\}\nonumber\\
 =&&\left\{(0, 0, 0, 0), \right. \nonumber\\
 &&(0, 0, -1, -1),(0, -1, 0, -1),(-1, 0, 0, -1),\nonumber\\
 &&(1, 0, 0, -1),(0, 1, 0, -1),(0, 0, 1, -1),\nonumber\\
 &&(0, 0, -1, 1),(0, -1, 0, 1),(-1, 0, 0, 1),\nonumber\\
 &&\left. (1, 0, 0, 1),(0, 1, 0, 1),(0, 0, 1, 1)\right\}.
\end{eqnarray}
At $i=1,...,12$, the function $s_h$ specifies the shifts $s=s_h(i)$ of multi-indices $n$, defined by the Fourier spectrum of the field $\textbf{A}$ (\ref{field}), which satisfy the condition $g_{4d}(s)=1$. Because of this, they are called the shifts of the first generation.
The sequential numbering $i=0,1,2,...$ of points $n=(n_1,n_2,n_3,n_4)=s_h(i)\in \mathcal{L}$, based on the use of $g_{4d}(n)$, takes into account the specific Fourier spectra of the electromagnetic lattice $\textbf{A}$ (\ref{field}) and the electron wave function $\Psi$~(\ref{sol1}) and thus drastically simplifies both numerical implementation of the presented approach and analysis of solutions~\cite{ESTCp2}.

We also use another useful numeration, namely, a specific numeration of 16 Dirac matrices $\Gamma_k, k=0,...,15$, which form a basis in the space of $4\times 4$ matrices~\cite{ESTCp1}. Any $4\times 4$ matrix $V=\sum_{k=0}^{15}V_k\Gamma_k$ is uniquely defined by the set $D_s(V)=\{V_k\}$ [Dirac set of matrix $V$ ($D$ set of $V$)]. The advantages of direct calculations with $D$ sets without matrix form retrieval are discussed in detail and illustrated in~\cite{ESTCp1,ESTCp3,ESTCp4}. Let us introduce the dimensionless parameters
\begin{equation}\label{Qw}
    \bm Q = (\textbf{q},i q_4) = {\bm K}/\kappa_e, \quad  \Omega = \frac{\hbar \omega_0}{m_e c^2},
\end{equation}
\begin{equation}\label{qq4}
\textbf{q} = q_1\textbf{e}_1+q_2\textbf{e}_2+q_3\textbf{e}_3 = \frac{\hbar \textbf{ k}}{m_e c}, \quad   q_4 = \frac{\hbar \omega}{m_e c^2}.
\end{equation}
In this notation, the matrix coefficients $V[n,s_h(i)]$ (\ref{meq}), in order of increasing $i=0,1,...,12$, have the following $D$ sets:
\begin{widetext}
\begin{eqnarray}
D_s\{V[n,(0,0,0,0)]\}&=&\{1,0,0,0,-w_4,0,0,0,0,0,0,0,0,i w_3,i w_1,i w_2\},\nonumber\\
D_s\{V[n,(0,0,-1,-1)]\}&=&\{0,0,0,0,0,0,0,0,0,0,0,0,0,0,-i a_{31}+b_{31},-i a_{32}+b_{32}\},\nonumber\\
D_s\{V[n,(0,-1,0,-1)]\}&=&\{0,0,0,0,0,0,0,0,0,0,0,0,0,-i a_{23}+b_{23},-i a_{21}+b_{21},0\},\nonumber\\
D_s\{V[n,(-1,0,0,-1)]\}&=&\{0,0,0,0,0,0,0,0,0,0,0,0,0,-i a_{13}+b_{13},0,-i a_{12}+b_{12}\},\nonumber\\
D_s\{V[n,(1,0,0,-1)]\}&=&\{0,0,0,0,0,0,0,0,0,0,0,0,0,-i a_{43}+b_{43},0,-i a_{42}+b_{42}\},\nonumber\\
D_s\{V[n,(0,1,0,-1)]\}&=&\{0,0,0,0,0,0,0,0,0,0,0,0,0,-i a_{53}+b_{53},-i a_{51}+b_{51},0\},\nonumber\\
D_s\{V[n,(0,0,1,-1)]\}&=&\{0,0,0,0,0,0,0,0,0,0,0,0,0,0,-i a_{61}+b_{61},-i a_{62}+b_{62}\},\nonumber\\
D_s\{V[n,(0,0,-1,1)]\}&=&\{0,0,0,0,0,0,0,0,0,0,0,0,0,0,-i a_{61}-b_{61},-i a_{62}-b_{62}\},\nonumber\\
D_s\{V[n,(0,-1,0,1)]\}&=&\{0,0,0,0,0,0,0,0,0,0,0,0,0,-i a_{53}-b_{53},-i a_{51}-b_{51},0\},\nonumber\\
D_s\{V[n,(-1,0,0,1)]\}&=&\{0,0,0,0,0,0,0,0,0,0,0,0,0,-i a_{43}-b_{43},0,-i a_{42}-b_{42}\},\nonumber\\
D_s\{V[n,(1,0,0,1)]\}&=&\{0,0,0,0,0,0,0,0,0,0,0,0,0,-i a_{13}-b_{13},0,-i a_{12}-b_{12}\},\nonumber\\
D_s\{V[n,(0,1,0,1)]\}&=&\{0,0,0,0,0,0,0,0,0,0,0,0,0,-i a_{23}-b_{23},-i a_{21}-b_{21},0\},\nonumber\\
D_s\{V[n,(0,0,1,1)]\}&=&\{0,0,0,0,0,0,0,0,0,0,0,0,0,0,-i a_{31}-b_{31},-i a_{32}-b_{32}\},\label{DsV}
\end{eqnarray}
\end{widetext}
where $n=(n_1,n_2,n_3,n_4), w_k=q_k+n_k \Omega, k=1,2,3,4$.

By taking into account Eqs.~(\ref{cn}) and (\ref{e14n})--(\ref{timejn}), the system of equations (\ref{meq}) with matrix coefficients $V(n,s)$ can be written in terms of scalar equations
\begin{eqnarray}\label{fC}
   \langle f^j(n),C\rangle &\equiv& \sum_{s \in S_{13}} V^j{}_k(n,s) c^k(n+s) = 0, \nonumber\\
  j&=&1,2,3,4;\quad  n \in {\mathcal L},
\end{eqnarray}
where
\begin{eqnarray}
  &&f^j(n)=\sum_{s \in S_{13}} V^j{}_k(n,s) \theta^k(n+s)\in V_C^\ast,\nonumber\\
  &&\langle f^j(n),e_k(n+s)\rangle=V^j{}_k(n,s).\label{fjn}
\end{eqnarray}
Finally, by combining the four equations related with each point $n$, one can rearrange Eqs.~(\ref{fC}) to the basic system of equations~\cite{ESTCp1,ESTCp4}
\begin{equation}\label{PnC}
    P(n)C = 0, \quad n \in {\mathcal L},
\end{equation}
where
\begin{equation}\label{Pnfaf}
    P(n) = [f^{\alpha}(n)]^\dag \otimes a^{\alpha}{}_\beta (n)f^\beta (n)
\end{equation}
is the Hermitian projection operator in $V_C$ with trace $tr[P(n)]=4$. The Hermitian $4\times 4$ matrices $a(n)$ are given in a explicit form in~\cite{ESTCp1,ESTCp4}.

Each amplitude $c(n)$ enters in 13 different matrix equations of the infinite system (\ref{meq}). This relatively simple structure of equations has made it possible to obtain the fundamental solution of the system~(\ref{PnC}) by a recurrent process~\cite{ESTCp1,ESTCp2,ESTCp4} based on a fractal approach~\cite{ESTCp2}. It is expressed in terms of an infinite series of projection operators. This process begins with the selection of an infinite subsystem consisting of independent equations and the calculation of the projection operators $\rho_0(n)=P(n), \quad n\in \mathcal{F}_0 \subset \mathcal{L}$, which uniquely define the fundamental solutions of these equations~\cite{ESTCp1,ESTCp4}. At each new $k$th step of the recurrent process, we add another infinite set $P(n)C = 0, n\in\mathcal{F}_k$ of mutually independent equations (MIE) which, however, are related with some of the equations introduced in the previous steps. Consequently, we obtain an infinite set of independent finite systems of interrelated equations [fractal clusters of equations (FCE)]. It can be described as a 4D lattice of such clusters. Each step of the recurrent procedure expands FCE for which it provides the exact fundamental solutions. The fractal algorithm of this expansion presented in \cite{ESTCp2}  is devised to minimize volumes of computations and data files. Some MIE (aggregative MIE, or MIE1) just add one equation to each cluster of the previous FCE lattice so that these enlarged clusters remain independent. Other MIE (connective MIE, or MIE2), by adding each equation, interrelate a pair of neighboring clusters into a joint cluster, and a quite different FCE lattice arises. Each fractal period includes connections in directions of $n_4, n_1, n_2$, and $n_3$ axes, respectively. The smaller the FCE, the smaller are the volumes of the computations and data files, which are necessary to find and to write the fundamental solution for this FCE. To simplify calculations, we add a maximal possible number of MIE1 before adding the next MIE2.

The fundamental solution  $\mathcal{S}$ of the system~(\ref{PnC}) is the Hermitian operator of projection onto the solution subspace of the multispinor space $V_C$. It is defined as follows~\cite{ESTCp1,ESTCp4}:
\begin{equation}\label{Pro}
  \mathcal{S}=\mathcal{U}-\mathcal{P},\quad  \mathcal{P}=\sum_{k=0}^{+\infty}\sum_{n\in\mathcal{F}_k}\rho_k(n),
\end{equation}
\begin{equation}\label{FkL}
    \bigcup_{k=0}^{+\infty}\mathcal{F}_k=\mathcal{L}, \quad \mathcal{F}_j\bigcap\mathcal{F}_k=\emptyset, \quad j\neq k,
\end{equation}
where $\rho_k(n)$ are Hermitian projection operators with trace $tr[\rho_k(n)]=4$. There exist various ways~\cite{ESTCp2} to split the lattice $\mathcal L$ into sublattices $\mathcal{F}_k$ to fulfill conditions (\ref{FkL}) and
\begin{eqnarray}\label{romn}
  \rho_k^{\dag}(n)&=&\rho_k^2(n)=\rho_k(n),\quad n \in \mathcal{L},\nonumber\\
  \rho_k(m)\rho_l(n)&=&0 \text{ if } k\neq l \text{ or (and) } m\neq n,\nonumber\\
  \rho_0(n)&=&P(n), \quad n\in \mathcal{F}_0,
\end{eqnarray}
which result in the relations $\mathcal{P}^\dag=\mathcal{P}^2=\mathcal{P}, \quad P(n)\mathcal{P}=\mathcal{P}P(n)=P(n)$, and, finally, $P(n)\mathcal{S}\equiv 0, n\in \mathcal{L}$. Hence, for any $C_0\in V_C$, $C=\mathcal{S}C_0$ is the exact particular solution of Eq.~(\ref{PnC}), specified by the multispinor $C_0$, i.e., the function $\Psi$~(\ref{sol1}) with the set of Fourier amplitudes $\{c(n), n \in \mathcal{L}\}=\mathcal{S}C_0$  satisfies the Dirac equation~(\ref{Dirac}) for the problem under consideration. Due to these properties, $\mathcal{P}$ is called the projection operator of the system of equations (\ref{PnC}). As shown in~\cite{ESTCp1}, this concept can be applied to any system of homogeneous linear equations.

It follows from Eq.~(\ref{Pnfaf}) that
\begin{equation}\label{PmPn}
    P(m)P(n)=\left[f^i(m)\right]^{\dag}\otimes\left[a(m)N(m,n)a(n)\right]^i{}_j f^j(n),
\end{equation}
where
\begin{equation}\label{Nij}
    N^i{}_j(m,n)=\left\langle f^i(m),\left[f^j(n)\right]^{\dag}\right\rangle,\quad i,j=1,2,3,4,
\end{equation}
$a(n) = [L(n)]^{-1}$, $L(n)\equiv  N(n,n)$, and $N(m,n)\equiv 0$ at $g_{4d}(n-m)>2$. Substitution of $f^{\alpha}(n)$ in (\ref{Nij}) at $n=m+s$ gives~ \cite{ESTCp1,ESTCp4}
\begin{eqnarray}\label{N1N2}
  N^{\dag}(n,m)=N(m,n)&=&L(m) \text{ for } n=m,\nonumber\\
                      &=&N_1(m,s) \text{ for } g_{4d}(s)=1,\nonumber\\
                      &=&N_2(s)U \text{ for } g_{4d}(s)=2,
\end{eqnarray}
where $U\equiv\Gamma_0$ is the $4\times 4$ unit matrix. The $D$ sets of 12 matrices $N_1(m,s)$ and the table of 56 scaler coefficients $N_2(s)$ for the general 4D-ESTC are presented in the Appendix. These major structural parameters of the ESTC specify interrelations in the system of equations~(\ref{PnC}). They are presented as functions of the dimensionless parameters $A_{jk}=a_{jk} + i b_{jk}, w_k=q_k+m_k\Omega$, and  $\Omega_{\pm}=\pm\Omega+2w_4$, where  $\Omega$ and $q_k$ are defined in Eqs.~(\ref{Qw}) and (\ref{qq4}), $k=1,2,3,4$, $m=(m_1,m_2,m_3,m_4)\in \mathcal{L}$.

The nonzero amplitudes for $\textbf{A}^\prime$~(\ref{Aprime}) are specified by $a_{12}=b_{13}=a_{42}=-b_{43}=A_m$. In this case, most of the structural parameters in Eq.~(\ref{N1N2})  are vanishing, only $N_1(m,s)$ with $D$ sets,
\begin{eqnarray}\label{Ds34}
  D_s&&\left\{N_1[m,(\mp 1,0,0,-1)]\right\}=\nonumber\\
     A_m &&\left\{2(-w_2 \mp i w_3),\mp i\Omega,0,-\Omega,0,0,0,0,\right.\nonumber\\\
     &&\left. 0, \mp i\Omega_{-},0,-\Omega_{-},0,0,0,0\right\},
\end{eqnarray}
\begin{eqnarray}\label{Ds910}
  D_s&&\left\{N_1[m,(\mp 1,0,0,1)]\right\}=\nonumber\\\
     A_m &&\left\{2(-w_2 \mp i w_3),\mp i\Omega,0,-\Omega,0,0,0,0,\right.\nonumber\\\
     &&\left. 0, \mp i\Omega_{+},0,-\Omega_{+},0,0,0,0\right\},
\end{eqnarray}
and $N_2(s)=4A_m^2$ with $s\in \{(0,0,0,-2),(0,0,0,2)\}$ are not zero.

\subsection{\label{sec:approx}Approximate particular solutions}
Numerical implementation of the obtained solution implies the replacement of the projection operator $\mathcal{P}$ (\ref{Pro}) of the infinite system of equations (\ref{PnC}) by the projection operator
\begin{equation}\label{Pprime}
    \mathcal{P'}=\sum_{k\in k_L}\sum_{n\in n_L(k)}\rho_k(n)
\end{equation}
of its finite subsystem
\begin{equation}\label{PkbLC}
    P(n)C=0,\quad n\in \mathcal{L}'=\bigcup_{k\in k_L}n_L(k) \subset \mathcal{L},
\end{equation}
where $k_L$ is an ordered finite list of integers, and $n_L(k)$ is a finite list of points $n\in\mathcal{F}_k$, taken into account. These lists define a finite model of the electron wave function in the ESTC, i.e., its approximation by a bispinor function with a finite discrete Fourier spectrum. Some such models are presented in~\cite{ESTCp2,ESTCp3,ESTCp4}. The projection operator
\begin{equation}\label{SkbL}
    \mathcal{S'} = \mathcal{U} - \mathcal{P'}
\end{equation}
gives the exact fundamental solution of the system~(\ref{PkbLC}), which is an approximate solution of the system~(\ref{PnC}).

Let $\mathcal{D}$ be a differential operator in a space $\mathcal{V}_{\Psi}$ of scalar, vector, spinor, or bispinor functions, and $\|\Psi\|$ be the norm of $\Psi$ on $\mathcal{V}_{\Psi}$. The functional
\begin{equation}\label{Rpsi}
    \mathcal{R}: \Psi\mapsto \mathcal{R}[\Psi]=\frac{\|\Psi_D\|}{\|\Psi\|}
\end{equation}
where $\Psi_D=\mathcal{D}\Psi$, evaluates the relative residual at the substitution of $\Psi$ into the differential equation $\mathcal{D}\Psi=0$. It provides a convenient fitness criterion to  accurately compare various approximate solutions of this equation~\cite{ESTCp2,ESTCp3,ESTCp4}. For an exact solution $\Psi$, the residual $\Psi_D$ vanishes, i.e., $\mathcal{R}[\Psi]=0$. If $\Psi_D\neq 0$, but $\mathcal{R}[\Psi]\ll 1$, the function $\Psi$ may be treated as a reasonable approximation to the exact solution, and the smaller is $\mathcal{R}[\Psi]$, the more accurate is the approximation. In terms of distances $d=\|\Psi\|$ and $d_D=\|\Psi_D\|$ of $\Psi$ and $\Psi_D$ to the origin of $\mathcal{V}_{\Psi}$ (the zero function), one can graphically describe $\mathcal{R}[\Psi]$  as shrinkage in distance $\mathcal{R}[\Psi]=d_D/d$. The functional $\mathcal{R}$, as applied to a family of functions $\Psi(\bm{x},\xi)$ with members specified by a parameter $\xi$, results in function $\mathcal{R}[\Psi(\bm{x},\xi)]$ of $\xi$, denoted $\mathcal{R}(\xi)$ for short.

In the present paper, $\mathcal{V}_{\Psi}=V_C$, the norm $\|\Psi\|$ is given by Eq.~(\ref{normpsi}), and $\Psi_D=\mathcal{D}\Psi$ is calculated for the dimensionless operator
\begin{equation}\label{Ddim}
    \mathcal{D}=\sum_{k=1}^3 \alpha_k\left(-\frac{i \hbar}{m_e c}\frac{\partial}{\partial x_k} - A'_k\right) - \frac{i \hbar}{m_e c^2}\frac{\partial}{\partial t} + \alpha_4
\end{equation}
of the equation~$\mathcal{D}\Psi=0$ equivalent to Eq.~(\ref{Dirac}). We restrict our consideration to the case when the amplitude $C_0$ specifying a particular solution is given by
\begin{equation}\label{C0a0no}
    C_0=a_0^j e_j(n_o),
\end{equation}
where $n_o=(0,0,0,0)$, and $q_2=q_3=0$, i.e., the electron moves along the axis $X_1$. The fitness parameter $\mathcal{R}(\xi)$ plays a leading role in search for the best approximate particular solution $\{c(n), n \in \mathcal{L}\}=\mathcal{S'}C_0$, available in the frame of the selected finite model, as follows.

The analytical fundamental solution $\mathcal{S}$~(\ref{Pro}) is obtained without recourse to any dispersion relation, i.e., for any vector $\bm Q$~(\ref{Qw}). However, since the system of equations~(\ref{PnC}) is homogeneous, the dispersion relation manifests itself in the spectral distribution of Fourier amplitudes $c(n)$ for each exact particular solution $\Psi$~(\ref{sol1}). This is illustrated in~\cite{ESTCp4} by the example of the exact Volkov solution. Since the amplitude $\Psi_0$~(\ref{sol1}) is periodic in $X_1, X_2, X_3$, and $X_4$, the wave function~$\Psi$ describes a nonlocalized solution of the Dirac equation. In the general case, its Fourier spectrum is also nonlocalized in the space of the four-dimensional wave vectors. However,  in numerical calculations for a finite model, instead of an exact particular solution, we obtain its approximation with a localized Fourier spectrum bounded by the truncation condition $ g_{4d}(n)\leq g_{max}$ for all $n\in\mathcal{L}'$. Consequently, the dispersion interrelation of $\textbf{q}$ and $q_4$ is defined by the minimum of the fitness function $\mathcal{R}=\mathcal{R}_1(\xi)$ with graphical representation in the form of a spectral curve of approximate solutions~\cite{ESTCp3,ESTCp4}, where
\begin{equation}\label{xiq4}
    \xi=q_4-\sqrt{1+\textbf{q}^2}=\frac{\hbar \omega}{m_e c^2}-\sqrt{1+\left(\frac{\hbar \textbf{k}}{m_e c}\right)^2}.
\end{equation}
Here, $\mathcal{R}_j=\sqrt{\lambda_j}$ is specified by a generalized eigenvalue $\lambda_j$ which is a root of the quartic equation $\det(U_D -\lambda U_E)=0$, with the Hermitian $4\times 4$ matrices $U_E$ and $U_D$, defined in~\cite{ESTCp2,ESTCp3,ESTCp4}. It has real coefficients and positive roots $\lambda_j$ indexed below in increasing order of magnitude, ${R}_1<{R}_2<{R}_3<{R}_4$. At sufficiently large value of $g_{max}$, the condition $\mathcal{R}_1\ll 1$ is satisfied within narrow limits of $\xi$ values, whereas $\mathcal{R}_{2,3,4}\gg\mathcal{R}_1$ and they do not satisfy the similar condition at any value of $\xi$; see numerical and graphic illustrations in~\cite{ESTCp3,ESTCp4}. The minimum  $\{\xi_0,\mathcal{R}_0=\mathcal{R}_1(\xi_0)\}$ of the curve $\mathcal{R}=\mathcal{R}_1(\xi)$ specifies the most accurate approximate solution provided by the selected finite model. The corresponding amplitude $a_0=a_{01}$~(\ref{C0a0no}) for this solution is specified by the generalized eigenvector $a_{01}$ defined by the equation $U_Da_{01}=\lambda_1 U_Ea_{01}$. It follows from the results of the computer simulations~\cite{ESTCp3,ESTCp4} that $\xi_0$ converges to a positive limit and $\mathcal{R}_0$ tends to zero with increasing $g_{max}$; in other words, this approximate particular solution converges to the exact solution with the dispersion relation $q_4-\sqrt{1+\textbf{q}^2}=\xi_0$.

\section{\label{sec:wave}Electron wave functions in the chiral 2d-ESTC}
\subsection{\label{sec:structure}Structure of wave functions}
For the problem under study the technique presented in~\cite{ESTCp1,ESTCp2,ESTCp3,ESTCp4} and Eqs.~(\ref{Ds34}) and (\ref{Ds910}) at $\textbf{q}=\textbf{q}_{\pm}\equiv\pm |q_1|\textbf{e}_1$ result in the four partial solutions
\begin{equation}\label{Psij}
    \Psi_j(\textbf{q}_{\pm})=\Psi_{j\pm} e^{i \Phi_{j\pm}},
\end{equation}
where $j=1, 2$ and
\begin{equation}\label{Phij}
    \Phi_{j\pm}=\bm{x}\cdot\bm{K}_{j\pm}=(\pm |q_1|\varphi_1-q_{4j}\varphi_4)/\Omega,
\end{equation}
\begin{eqnarray}\label{Psi1pm2pm}
  \Psi_{1\pm}&=&\sum_{n \in \mathcal S_{1\pm}} a_{\pm}(n) e^{i \varphi_n},\quad
  \Psi_{2\pm}=\sum_{n \in \mathcal S_{2\pm}} b_{\pm}(n) e^{i \varphi_n},\nonumber \\
  \varphi_n&=&n_1 \varphi_1-n_4 \varphi_4,\quad \varphi_1=2\pi X_1, \varphi_4=n_4 X_4.
\end{eqnarray}
The points $n=(n_1,n_2,n_3,n_4)\in \mathcal L$ with nonzero bispinor Fourier amplitudes $a_{\pm}(n)$ and $b_{\pm}(n)$, comprising the solution domains $S_{1\pm}$ and $S_{2\pm}$, satisfy the conditions $|n_1|=0, 1; n_2=n_3=0$. These amplitudes, calculated by the recurrent algorithm~\cite{ESTCp1}, have specific symmetry properties which make it possible to express $\Psi_{1\pm}$ and $\Psi_{2\pm}$ in terms of eight complex scalar functions $z_{jk}=z_{jk}(\varphi_4)$ as follows:
\begin{eqnarray}\label{Psi12pm}
  \Psi_{1+}&=&u_2 z_{12}+u_4 z_{14}+i e^{i\varphi_1}(u_1 z_{11}+ u_3 z_{13}),\nonumber\\
  \Psi_{1-}&=-&u_1 z_{12}-u_3 z_{14}+i e^{-i\varphi_1}(u_2 z_{11}+ u_4 z_{13}),\nonumber\\
  \Psi_{2+}&=&u_1 z_{21}+u_3 z_{23}+i e^{-i\varphi_1}(u_2 z_{22}+ u_4 z_{24}),\nonumber\\
  \Psi_{2-}&=-&u_2 z_{21}-u_4 z_{23}+i e^{i\varphi_1}(u_1 z_{22}+ u_3 z_{24}),
\end{eqnarray}
where
\begin{eqnarray}\label{u1234}
  u_1&=&\frac{1}{\sqrt{2}}\left(\begin{array}{c}
          1 \\
          1 \\
          0 \\
          0
        \end{array}\right),
         u_2=\frac{1}{\sqrt{2}}\left(\begin{array}{c}
          1 \\
          -1 \\
          0 \\
          0
        \end{array}\right),\nonumber\\
  u_3&=&\frac{1}{\sqrt{2}}\left(\begin{array}{c}
          0 \\
          0 \\
          1 \\
          1
        \end{array}\right),
         u_4=\frac{1}{\sqrt{2}}\left(\begin{array}{c}
          0 \\
          0 \\
          1 \\
          -1
        \end{array}\right).
\end{eqnarray}

The interrelations between the complex scalar functions $z_{jk}$ and the bispinor amplitudes $a_{\pm}(n), b_{\pm}(n)$ are described by the Fourier expansions
\begin{equation}\label{Zexp}
    \mathcal{Z}_j=\sum_{l=-\infty}^{+\infty} \mathcal{Z}_{j,l} e^{il\varphi_4},
\end{equation}
where $j=1, 2,$
\begin{equation}\label{ZmZp}
    \mathcal{Z}_{1}=\left(\begin{array}{c}
                  z_{12}\\
                  z_{14}\\
                  z_{11}\\
                  z_{13}
                \end{array}\right), \mathcal{Z}_{2}=\left(\begin{array}{c}
                  z_{21}\\
                  z_{23}\\
                  z_{22}\\
                  z_{24}
                \end{array}\right),
\end{equation}
\begin{equation}\label{Zeven}
    \mathcal{Z}_{1, l}=\left(\begin{array}{c}
                  a_{-l2}\\
                  a_{-l4}\\
                  0\\
                  0
                \end{array}\right), \mathcal{Z}_{2, l}=\left(\begin{array}{c}
                  b_{-l1}\\
                  b_{-l3}\\
                  0\\
                  0
                \end{array}\right),
\end{equation}
\begin{eqnarray}\label{abeven}
  a_{+}[(0,0,0,l)]&=&a_{l2}u_2+a_{l4}u_4,\nonumber\\
  b_{+}[(0,0,0,l)]&=&b_{l1}u_1+b_{l3}u_3
\end{eqnarray}
for even $l$, and
\begin{equation}\label{Zodd}
    \mathcal{Z}_{1, l}=\left(\begin{array}{c}
                  0\\
                  0\\
                  a_{-l1}\\
                  a_{-l3}
                \end{array}\right), \mathcal{Z}_{2, l}=\left(\begin{array}{c}
                  0\\
                  0\\
                  b_{-l2}\\
                  b_{-l4}
                \end{array}\right),
\end{equation}
\begin{eqnarray}\label{abodd}
  a_{+}[(1,0,0,l)]&=&i (a_{l1}u_1+a_{l3}u_3),\nonumber\\
  b_{+}[(-1,0,0,l)]&=&i (b_{l2}u_2+b_{l4}u_4)
\end{eqnarray}
for odd $l$. All scalar coefficients $a_{lk}$ and  $b_{lk}$ are real. In accordance with the above definitions, the functions $z_{jk}$ are given by $a_{+}(n), b_{+}(n)$. However, they also specify $\Psi_{1-}$ and $\Psi_{2-}$ through the relations between $a_{-}(n), b_{-}(n)$ and $a_{+}(n), b_{+}(n)$, taken into account in Eq.~(\ref{Psi12pm}).

The bispinor functions $\Psi_j(\textbf{q}_{\pm})$ are uniquely defined by eight complex scalar functions (structural functions) $z_{jk} (j=1,2; k=1,2,3,4)$, which serve as convenient building blocks of the relations describing the electron properties. For the chiral 2D-ESTC under study, the finite model [see Eq.~(\ref{PkbLC})] is given by $\mathcal{L}'=\{n=(n_1,0,0,n_4), 0\leq g_{4d}(n)\leq g_{max}\}$, i.e., the infinite series in Eq.~(\ref{Zexp}) are truncated so that the real $x_{jk}$ and imaginary $y_{jk}$ parts of $z_{jk}=x_{jk}+i y_{jk}$ can be written as
\begin{eqnarray}\label{xy1223}
  x_{jk}&=&x_{jk0}+\sum_{p=1}^{p_m}x_{jk(2p)}\cos(2p \varphi_4),\nonumber\\
  y_{jk}&=&\sum_{p=1}^{p_m}y_{jk(2p)}\sin(2p \varphi_4)
\end{eqnarray}
for $jk\in \{12,14,21,23\}$, and
\begin{eqnarray}\label{xy1124}
  x_{jk}&=&\sum_{p=0}^{p_m}x_{jk(2p+1)}\cos[(2p+1) \varphi_4],\nonumber\\
  y_{jk}&=&\sum_{p=0}^{p_m}y_{jk(2p+1)}\sin[(2p+1) \varphi_4]
\end{eqnarray}
for $jk\in \{11,13,22,24\}$, where
\begin{eqnarray}\label{xyab}
  x_{1k0}&=&a_{0k},\quad x_{2k0}=b_{0k},\nonumber \\
  x_{1kl}&=&a_{-lk}+a_{lk},\quad  x_{2kl}=b_{-lk}+b_{lk},\nonumber\\
  y_{1kl}&=&a_{-lk}-a_{lk},\quad  y_{2kl}=b_{-lk}-b_{lk}.
\end{eqnarray}
By selecting a sufficiently large value of $g_{max}$, one can easily obtain approximate solutions with any desired accuracy. To illustrate this, we fix $\Omega=0.01$, $A_m=\sqrt{2}/200$ and set $g_{max}=12$ for which $p_m=6$. This results in the approximate particular solutions satisfying the fitness condition $\mathcal{R}_0<10^{-17}$, presented below, whose deviations from the corresponding exact solutions are negligibly small.

\subsection{\label{sec:dispersion}Dispersion relations}
\begin{figure}
\includegraphics{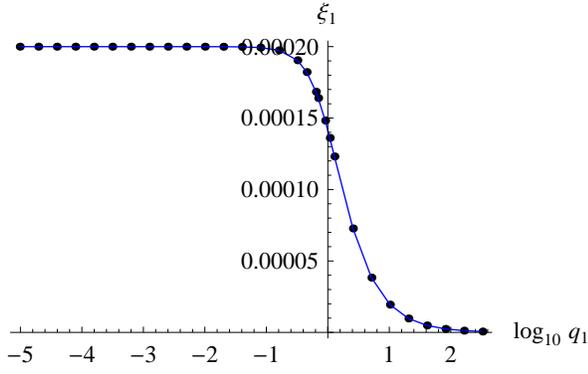}
\caption{\label{fig1x0a}Plot of $\xi_1$ against $\log_{10}q_1$.}
\end{figure}

\begin{figure}
\includegraphics{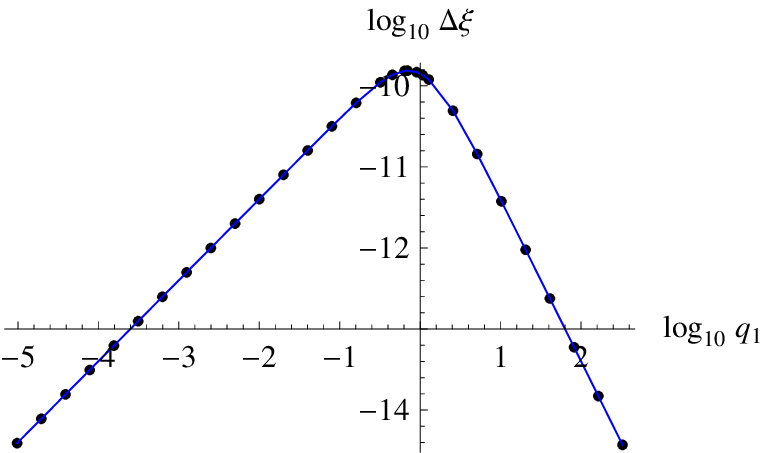}
\caption{\label{fig2dx0}Plot of $\Delta\xi$ against $\log_{10}q_1$.}
\end{figure}

For a given $q_1\neq 0$, the dispersion equation has two closely spaced solutions $q_{4j}=q_{40}+\xi_j$, where $j=1, 2, \xi_1<\xi_2$ and $q_{40}=\sqrt{1+q_1^2}$. They are invariant under inversion $q_1\mapsto -q_1$. The electron wave functions $\Psi_j(\textbf{q}_{+})$ and $\Psi_j(\textbf{q}_{-})$ describe the motion in the positive and negative $X_1$ directions, respectively. The dependence of $\xi_1$ and $\Delta \xi =\xi_2-\xi_1$ on $q_1$ is shown in Figs.~\ref{fig1x0a} and \ref{fig2dx0}, where the dots represent calculations at $q_1= 2^m \Omega, m\in [-10, 15]$, while the curves are obtained by the linear interpolation.

\subsection{\label{sec:zjk}Properties of functions $z_{jk}$}
Substitution of $\Psi_j(\textbf{q}_{+})$ in the Dirac equation $\mathcal{D}\Psi =0$ with~$\mathcal{D}$~(\ref{Ddim}) result in two evolution equations,
\begin{equation}\label{dZj}
    \frac{\rm d}{{\rm d}\varphi_4}\mathcal{Z}_j=\frac{i}{\Omega}M_j\mathcal{Z}_j,\quad j=1, 2,
\end{equation}
where
\begin{equation}\label{Mj}
    M_j=N_j - (-1)^j 4A_m\cos{\varphi_4} \alpha_1,
\end{equation}
\begin{equation}\label{Mm}
    N_j=\left(
            \begin{array}{cccc}
              q_{4j}-1&-q_{1j}&0&0\\
              -q_{1j}&q_{4j}+1&0&0\\
              0&0&q_{41}-1&q_{1j}-\Omega\\
              0&0&q_{1j}-\Omega&q_{4j}+1\\
            \end{array}
          \right),
\end{equation}
\begin{equation}\label{alpha1}
    \alpha_1=\left(
            \begin{array}{cccc}
              0&0&0&1\\
              0&0&1&0\\
              0&1&0&0\\
              1&0&0&0\\
            \end{array}
          \right), \quad q_{1j}=(-1)^j q_1.
\end{equation}

Since $M_1$ and $M_2$ are real symmetrical matrices, it follows from Eqs.~(\ref{dZj}) that ${\rm d}(\mathcal{Z}^{\dag}_j\mathcal{Z}_j)/{\rm d}\varphi_4=0$. Therefore, we impose the normalization condition
\begin{equation}\label{normZj}
    \mathcal{Z}^{\dag}_j\mathcal{Z}_j\equiv \Psi_{j\pm}^\dag\Psi_{j\pm}\equiv\sum_{k=1}^4 |z_{jk}|^2=1, j=1, 2.
\end{equation}
The functions $\Psi_{1\pm}$ and $\Psi_{2\pm}$ also satisfy the following orthonormality relations:
\begin{eqnarray}
  \Psi_{j\pm}^\dag\Psi_{j\mp}&=&0,\quad j=1,2,\nonumber \\
  \Psi_{1\pm}^\dag\Psi_{2\pm}&=&0,\quad \lim_{q_1\rightarrow 0}(\Psi_{1\pm}^\dag\Psi_{2\mp})=1.
\end{eqnarray}
In our numerical calculations with $g_{max}=12$, variations from these relations are negligibly small, at less than $10^{-16}.$

\begin{figure}
\includegraphics{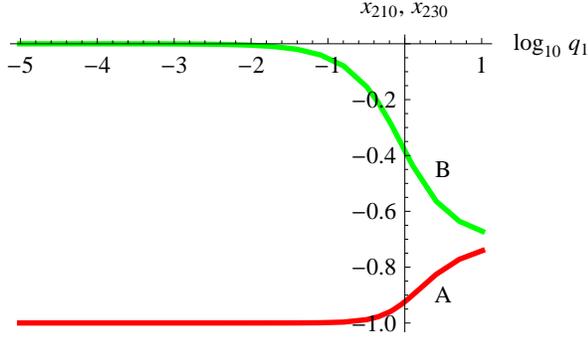}
\caption{\label{fig3x210x230}Plot of (A) $x_{210}$ and (B) $x_{230}$  against $\log_{10}q_1$.}
\end{figure}

\begin{figure}
\includegraphics{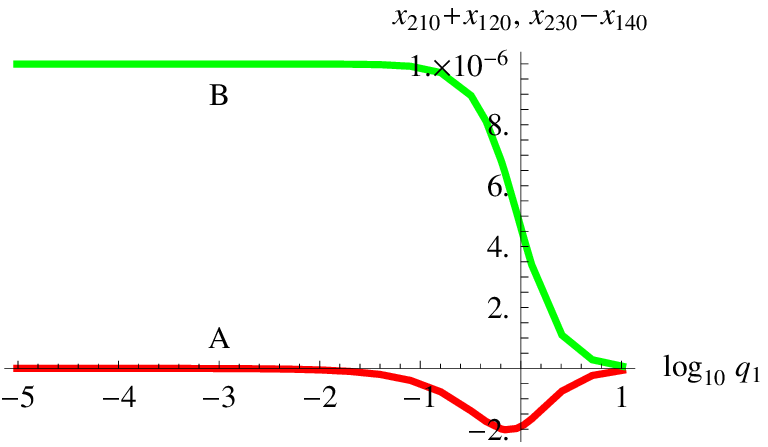}
\caption{\label{fig4dx}Plot of (A) $x_{210}+x_{120}$ and (B) $x_{230}-x_{140}$  against $\log_{10}q_1$.}
\end{figure}

\begin{figure}
\includegraphics{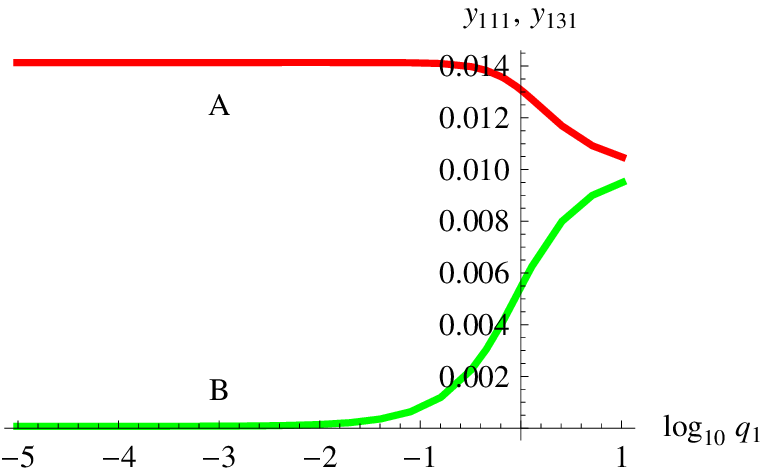}
\caption{\label{fig5y111y131}Plot of (A) $y_{111}$ and (B) $y_{131}$  against $\log_{10}q_1$.}
\end{figure}

\begin{figure}
\includegraphics{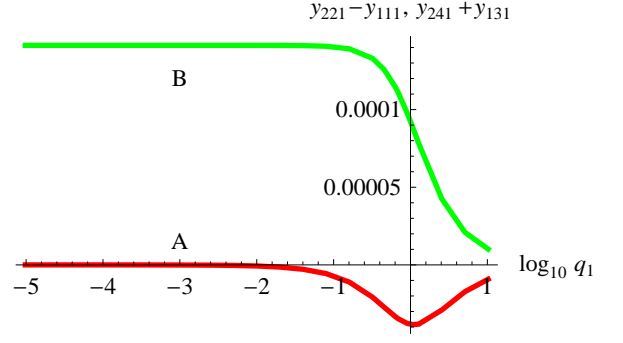}
\caption{\label{fig6dy}Plot of (A) $y_{221}-y_{111}$ and (B) $y_{241}+y_{131}$  against $\log_{10}q_1$.}
\end{figure}

Substitution of $\mathcal{Z}_1$ and $\mathcal{Z}_2$ into Eqs.~(\ref{dZj}) results in two independent systems of matrix equations in Fourier amplitudes $\mathcal{Z}_{1, l}$ and $\mathcal{Z}_{2,l}$, respectively,
\begin{equation}\label{Zeq}
    (N_j-l\Omega U)\mathcal{Z}_{j, l}=(-1)^j 2 A_m\alpha_1(\mathcal{Z}_{j, l-1}+\mathcal{Z}_{j, l+1}),
\end{equation}
where $j=1, 2$. These amplitudes are connected by the recurrent relations
\begin{equation}\label{Zplus}
    \mathcal{Z}_{j, l+1}=-\mathcal{Z}_{j, l-1} + \frac{(-1)^j}{2A_m}\alpha_1 (N_j-l \Omega U) \mathcal{Z}_{j, l}
\end{equation}
with $l=1, 2,...$, and
\begin{equation}\label{Zminus}
    \mathcal{Z}_{j, l-1}=-\mathcal{Z}_{j, l+1} + \frac{(-1)^j}{2A_m}\alpha_1 (N_j-l \Omega U) \mathcal{Z}_{j, l}
\end{equation}
with $l=-1, -2,...$, where
\begin{equation}\label{Zeq0}
    N_j\mathcal{Z}_{j, 0}=(-1)^j 2 A_m\alpha_1(\mathcal{Z}_{j, -1}+\mathcal{Z}_{j, 1}).
\end{equation}
Therefore, by taking into account Eqs.~(\ref{Zeven}), (\ref{Zodd}), and (\ref{xyab}), coefficients $x_{jkl}, y_{jkl}$ can be calculated starting with $x_{120}, x_{140}, y_{111}, y_{131}$ and $x_{210}, x_{230}, y_{221}, y_{241}$. These starting coefficients  depend on $q_1$, as illustrated in Figs.~\ref{fig3x210x230}--\ref{fig6dy}.

In the state defined by the quasimomentum $\textbf{p}=\hbar\textbf{k}=m_e c\textbf{q}=0$, the equation $(U_D -\lambda_1 U_E)a_0=0$ has the twofold generalized eigenvalue $\lambda_1$ and the two-dimensional subspace of the corresponding generalized eigenvectors $a_0$. Any basis of this subspace specifies two linearly independent solutions of the Dirac equation, for which $\xi_1=\xi_2=0.000199970011$. In particular, the limiting cases of $\Psi_j(\textbf{q}_{\pm})$~(\ref{Psij}) at $q_1\rightarrow 0$ can be conveniently treated as the basis wave functions. At $q_1=0$, the functions $z_{jk}$ satisfy the identities
\begin{equation}\label{zequal}
    z_{11}=z_{22}, z_{12}+z_{21}=0, z_{13}=z_{24}, z_{14}+z_{23}=0
\end{equation}
and hence
\begin{eqnarray}\label{limpsi}
  \lim_{q_1\rightarrow 0}\Psi_{1\pm}&=&\lim_{q_1\rightarrow 0}\Psi_{2\mp},\nonumber\\
  \lim_{q_1\rightarrow 0}\Psi_1(\textbf{q}_{\pm})&=&\lim_{q_1\rightarrow 0}\Psi_2(\textbf{q}_{\mp}).
\end{eqnarray}
The coefficients, illustrated in Figs.~\ref{fig3x210x230}--\ref{fig6dy},  at this state have the following values:
\begin{eqnarray}
 x_{120}&=&-x_{210}=0.999875,\nonumber\\
 x_{140}&=&-x_{230}=-4.99594\times 10^{-7},\nonumber\\
 y_{111}&=&y_{221}=0.0141368,\nonumber\\
 y_{131}&=&y_{241}=0.0000706745.
\end{eqnarray}

\subsection{\label{sec:birefri}Energy level splitting}
Let us now compare the wave functions $\Psi_j(\textbf{q}_{\pm})$, $j=1, 2$ in terms of the corresponding mean values of Hamiltonian
\begin{equation}\label{Hamilton}
    H=c\sum_{k=1}^3\alpha_k p_k + m_e c^2\alpha_4,
\end{equation}
operators of kinetic momentum
\begin{equation}\label{pk}
   p_k=-i\hbar\frac{\partial}{\partial x_k} - \frac{e}{c}A_k,
\end{equation}
probability current density (velocity)  $j_k=c\alpha_k$, and spin $S_k=\frac{\hbar}{2}\Sigma_k$, $k=1, 2, 3$. Since $\Psi_j^\dag(\textbf{q}_{\pm})\Psi_j(\textbf{q}_{\pm})=1$, the mean value $\langle L \rangle$ of a linear operator $L$ with respect to the wave function $\Psi_j(\textbf{q}_{\pm})$ reduces to the mean value of the corresponding Hermitian form: $\langle L \rangle=\langle\Psi_j^\dag(\textbf{q}_{\pm}) L \Psi_j(\textbf{q}_{\pm})\rangle $. The mean values $\langle j_k \rangle$, $\langle p_k \rangle$, and $\langle S_k \rangle$ are zero at $k=2, 3$ for all these functions. For both $\Psi_1(\textbf{q}_{\pm})$ and $\Psi_2(\textbf{q}_{\pm})$, the inversion $\textbf{q}_{\pm}\mapsto\textbf{q}_{\mp}$ changes the signs of $\langle j_1 \rangle$, $\langle p_1 \rangle$, and $\langle S_1 \rangle$, but leaves invariant $\langle H \rangle$. It also follows from the results of our calculations that
\begin{equation}\label{meansigma1}
    \Psi_j^\dag(\textbf{q}_{\pm}) \Sigma_1 \Psi_j(\textbf{q}_{\pm})\equiv\Psi_{j\pm}^\dag \Sigma_1 \Psi_{j\pm}=\pm (-1)^j \Sigma_{10},
\end{equation}
where $\Sigma_{10}$ can be expressed in terms of the functions $z_{jk}$ as
\begin{equation}\label{sigma0}
    \Sigma_{10}= (-1)^j (|z_{j1}|^2+|z_{j3}|^2-|z_{j2}|^2-|z_{j4}|^2), j=1, 2.
\end{equation}
It is independent of $q_1$ and for the chiral ESTC under consideration takes the value $\Sigma_{10}=0.99960023984$.

The functions  $\Psi_1(\textbf{q}_{-})$ and $\Psi_2(\textbf{q}_{+})$ provide the same positive mean value ${\langle S_1\rangle}=S_{+}=\frac{\hbar}{2}\Sigma_{10}$, whereas $\Psi_1(\textbf{q}_{+})$ and $\Psi_2(\textbf{q}_{-})$ provide the same negative mean value ${\langle S_1\rangle}=S_{-}=-\frac{\hbar}{2}\Sigma_{10}$. Hence, $\Psi_1(\textbf{q}_{\mp})$ together with $\Psi_2(\textbf{q}_{\pm})$ specify two wave functions ($S_{\pm}$ solutions) describing two different spin states and defined into the whole united $q_1$ domain containing both $q_1<0$ and $q_1\geq 0$ values.

Let $J_{1\pm}(q_1)=\langle j_1 \rangle/c$, $P_{1\pm}(q_1)=\langle p_1 \rangle/(m_{e}c)$, and  $E_{\pm}(q_1)=\langle H \rangle/(m_{e}c^2)$ be the normalized mean values of the operators $j_1$, $p_1$, and $H$ with respect to $S_{\pm}$ solutions at a given $q_1$. At $q_1=0$, these solutions provide the same value of the normalized energy  $E_0=E_{\pm}(0)=1.000199970009$, and equal in magnitude but opposite in sign the normalized mean values of the velocity $J_{1\pm}(0)=\pm v_{10}$  and the momentum $P_{1\pm}(0)=\mp p_{10}$, where $v_{10}=1.99820142893 \times 10^{-10}$ and $p_{10}=1.99880079944\times 10^{-6}$.

\begin{figure}
\includegraphics{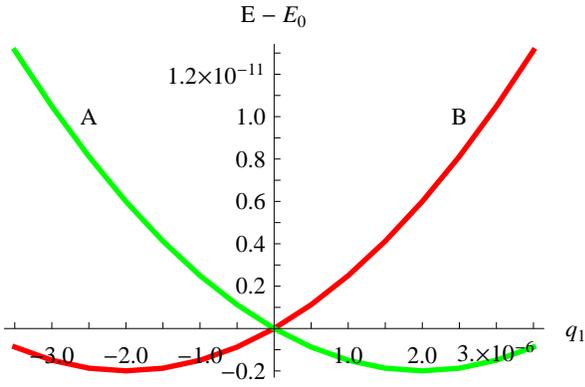}
\caption{\label{fig7hq0}Difference of the normalized energy $E=E_{\pm}(q_1)$ and $E_0$ against $q_1$:  (A) $E=E_{+}(q_1)$ and (B) $E=E_{-}(q_1)$.}
\end{figure}

\begin{figure}
\includegraphics{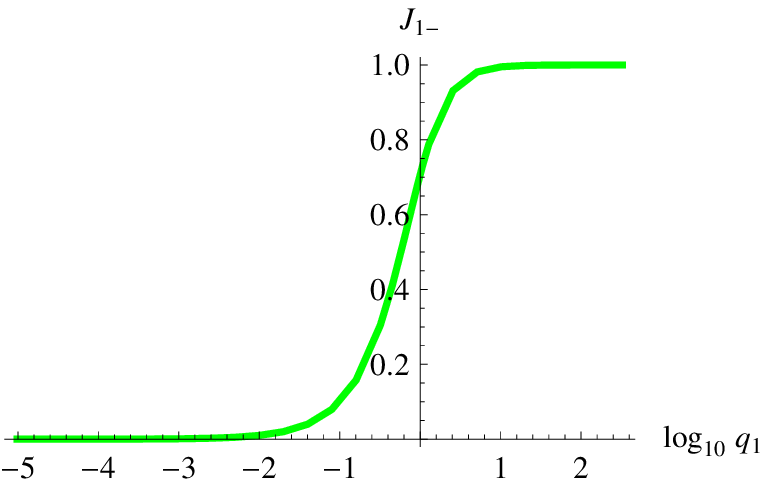}
\caption{\label{fig8J1m}Plot of $J_{1-}$ against $\log_{10}q_1$.}
\end{figure}

\begin{figure}
\includegraphics{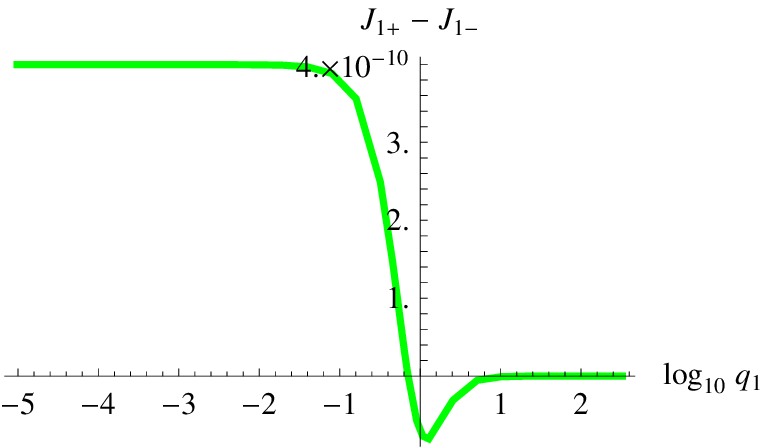}
\caption{\label{fig9dJ1}Plot of $J_{1+}-J_{1-}$ against $\log_{10}q_1$.}
\end{figure}

\begin{figure}
\includegraphics{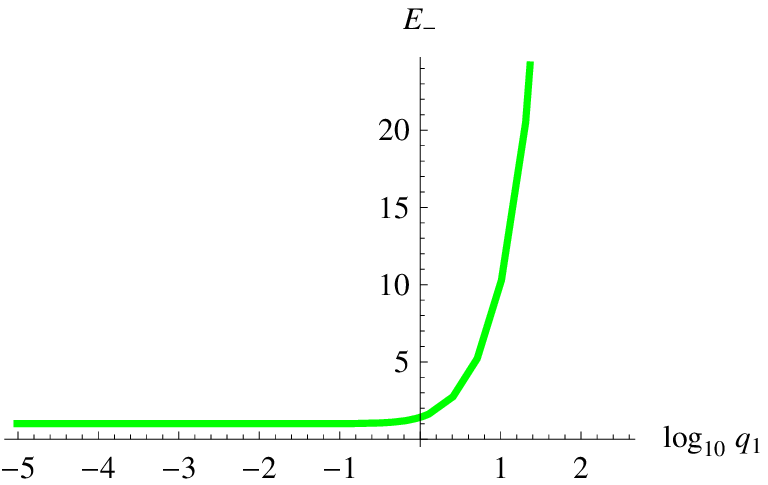}
\caption{\label{fig10Em}Normalized energy $E_{-}$ against $\log_{10}q_1$.}
\end{figure}

\begin{figure}
\includegraphics{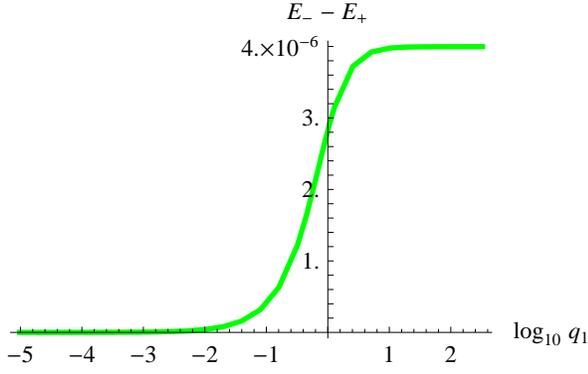}
\caption{\label{fig11dE}Normalized energy difference $E_{-}-E_{+}$ against $\log_{10}q_1$.}
\end{figure}

The mean values of momentum for $S_{\pm}$ solutions linearly depend on the quasimomentum: $P_{1\pm}(q_1)=q_1 \mp p_{10}$. The dependence of $E_{\pm}$ on $q_1$ in the vicinity of the origin is shown in Fig.~\ref{fig7hq0}. In this domain, $J_{1\pm}(q_1)$ can be closely approximated by the linear functions with  $J_{1\pm}(q_1)=0$ at $q_1=\mp q_{10}$, respectively, where $q_{10}=1.99860096936 \times 10^{-10}$.   At $|q_1|< q_{10}$, the mean values $J_{1\pm}(q_1)$ and $P_{1\pm}(q_1)$ are opposite in sign for both of the solutions. Figures~\ref{fig8J1m}--\ref{fig11dE} illustrate the properties of functions $J_{1\pm}(q_1)$ and $E_{\pm}(q_1)$ over a wide range of $q_1>0$. At any $q_1\neq 0$, there are two different states with the opposite in sign spins $S_{\pm}$ and different energy levels $E_{\pm}(q_1)$. This energy level splitting satisfies the relations $E_{-}(q_1)-E_{+}(q_1)=E_{+}(-q_1)-E_{-}(-q_1)\geq 0$ for $q_1\geq 0$. The functions $E_{\pm}(q_1)$ take the same minimal value $E_{min}=E_{\pm}(\pm p_{10})=1.000199970007$ at the points $q_1=\pm p_{10}$, where $P_{1\pm}(\pm p_{10})=0$. The wave functions $\Psi_2(\textbf{q}_{\pm})$ specify these two ground states with oppositely directed spins.

\section{\label{sec:family}Unidirectional and bidirectional states of the electron}
At $q_1\neq 0$, the wave functions $\Psi_j(\textbf{q}_{\pm}), j=1, 2$ are linearly independent and form a basis for a  four-dimensional subspace of partial solutions to the Dirac equation. At $q_1=0$, as a consequence of Eq.~(\ref{limpsi}), this subspace degenerates to the two-dimensional one. In this section, we treat  two families of partial solutions which describe unidirectional and bidirectional states of the Dirac electron. They are specified by the wave functions
\begin{eqnarray}\label{unibi}
  \Psi_{\pm}&=&\Psi_1(\textbf{q}_{\pm})e^{i\delta}\cos\alpha + \Psi_2(\textbf{q}_{\pm})\sin\alpha,\nonumber\\
  \Psi_j&=&\Psi_j(\textbf{q}_{+})e^{i\delta}\cos\alpha + \Psi_j(\textbf{q}_{-})\sin\alpha, j=1,2,
\end{eqnarray}
where $\alpha\in [0,\pi/2]$ and $\delta\in [0,2\pi]$. To study these states, we use the structural functions $z_{jk}$ described in Sec.~\ref{sec:structure} and the following designations:
\begin{eqnarray}
  R_{ikjl}&=&2\rm{Re}(z_{ik}^{\ast}z_{jl}),\quad I_{ikjl}=2\rm{Im}(z_{ik}^{\ast}z_{jl}),\nonumber\\
  R_{j}&=&R_{j1j4} + R_{j2j3},\quad I_{j}=I_{j1j2} + I_{j3j4},\nonumber\\
  v_{1j}&=&R_{j1j3}-R_{j2j4},
\end{eqnarray}
where $i, j=1, 2$ and $k, l=1, 2, 3, 4$.

\subsection{\label{sec:unidir}Unidirectional states: precession}
The wave functions $\Psi_{+}$ and $\Psi_{-}$ describe various electron states specified by the parameters $\alpha$ and $\delta$ at the positive quasimomentum $q_1$ and the negative one, respectively. For any linear operator $A$, the Hermitian forms $\Psi_{\pm}^\dag A \Psi_{\pm}$ are periodic in $X_1$ with the unit period. They are not periodic in $X_4$ because of the phase difference,
\begin{equation}\label{uniphi}
    \varphi=\Phi_{1\pm}-\Phi_{2\pm}+\delta=\frac{2\pi}{\Omega}\Delta\xi X_4 + \delta.
\end{equation}
However, $\Delta\xi/\Omega\ll 1$, so that variations of $\varphi$ at any unit interval of the $X_4$ axis are negligibly small in the calculation of norms and mean values. In this approximation, one can obtain the relations
\begin{eqnarray}\label{uniP1E}
  P_{1\pm}={\mathcal I}_{\Delta X}(\Psi_{\pm}^\dag p_1 \Psi_{\pm})/(m_{e}c)&=&\pm (|q_1|+ p_{10}\cos 2\alpha),\nonumber\\
  E={\mathcal I}_{\Delta X}(\Psi_{\pm}^\dag H \Psi_{\pm})/(m_{e}c^2)&=&E_1\cos^2\alpha+E_2\sin^2\alpha,\nonumber\\
  \Psi_{\pm}^\dag\Psi_{\pm}&=&1,
\end{eqnarray}
where
\begin{equation}\label{p10}
    p_{10}=\frac{\Omega}{2}(1-\Sigma_{10}),\quad  E_j=\int_0^1 H_j d X_4, j=1, 2,
\end{equation}
\begin{eqnarray}\label{uniEj}
  H_j&=&\Omega R_{jjjj+2} + |q_1| v_{1j}+|z_{j1}|^2 + |z_{j2}|^2\nonumber\\
  &-&|z_{j3}|^2 -|z_{j4}|^2+(-1)^j 4 A_m R_{j}\cos\varphi_4,
\end{eqnarray}
\begin{equation}\label{IntX}
    {\mathcal I}_{\Delta X}(f)\equiv\int_{\Delta X}f dX_1dX_4,
\end{equation}
and the domain $\Delta X$ is given by the unit intervals $[X_k,X_k+1], k=1, 4$. The dependence of the normalized energies $E_1=E_{-}(|q_1|)$ and $E_2=E_{+}(|q_1|)$ on $q_1$ is illustrated in Figs.~\ref{fig7hq0}, \ref{fig10Em}, and \ref{fig11dE}.

\begin{figure}
\includegraphics{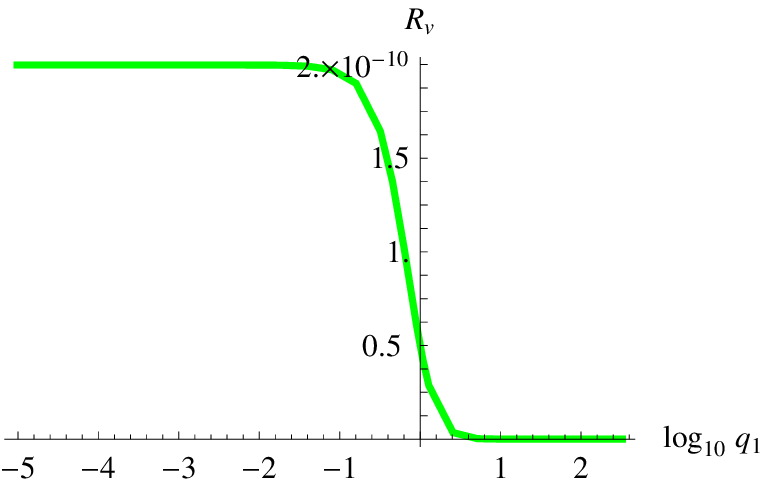}
\caption{\label{fig12Rv}Precession parameter $R_v$ against $\log_{10}q_1$.}
\end{figure}

\begin{figure}
\includegraphics{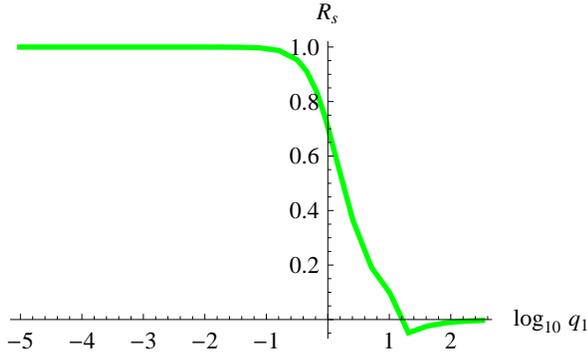}
\caption{\label{fig13Rs}Precession parameter $R_s$ against $\log_{10}q_1$.}
\end{figure}

In the comparative analysis of electron states, it is advantageous to calculate both mean values and Hermitian forms of various operators with respect to the corresponding wave functions. In particular, the Hermitian forms for the velocity operator and the spin operator with respect to $\Psi_{\pm}$ result in the following vector fields:
\begin{equation}\label{unijS}
    \textbf{j}(\textbf{q}_{\pm})=c\textbf{v}_{\pm},\quad \textbf{S}(\textbf{q}_{\pm})=\frac{\hbar}{2}\textbf{s}_{\pm},
\end{equation}
where
\begin{eqnarray}\label{univpm}
\textbf{v}_{\pm}&=&\sum_{k=1}^3 \textbf{e}_k (\Psi_{\pm}^\dag \alpha_k\Psi_{\pm})=\textbf{e}_1 \left\{\pm (v_{11}\cos^2\alpha + v_{12}\sin^2\alpha) \right.\nonumber\\
&+& \left. \sin 2\alpha \left[ \rm{Im}\,C_1 \cos(\varphi \pm\varphi_1) - \rm{Re}\,C_1 \sin(\varphi \pm\varphi_1) \right]\right\}\nonumber \\
&+& (R_{1}\cos^2\alpha - R_{2}\sin^2\alpha )\textbf{e}_A(\varphi_1) \nonumber\\
&+&\sin 2\alpha\left[\pm\rm{Im}\,C_2 \textbf{e}_A(\mp\varphi) +\rm{Re}\,C_2 \textbf{e}_B(\mp\varphi) \right.\nonumber\\
&\pm&\left.\rm{Im}\,C_3 \textbf{e}_A(2\varphi_1\pm\varphi) +\rm{Re}\,C_3 \textbf{e}_B(2\varphi_1\pm\varphi)\right],
\end{eqnarray}
\begin{eqnarray}\label{unispm}
\textbf{s}_{\pm}&=&\sum_{k=1}^3 \textbf{e}_k (\Psi_{\pm}^\dag \Sigma_k\Psi_{\pm})=\textbf{e}_1 \left\{\mp \Sigma_{10}\cos 2\alpha \right.\nonumber\\
&+& \left. \sin 2\alpha \left[ \rm{Im}\,D_1 \cos(\varphi \pm\varphi_1) - \rm{Re}\,D_1 \sin(\varphi \pm\varphi_1) \right]\right\}\nonumber \\
&\pm&(I_{1}\cos^2\alpha - I_{2}\sin^2\alpha )\textbf{e}_B(\varphi_1)\nonumber \\
&+&\sin 2\alpha\left[\pm\rm{Im}\,D_2 \textbf{e}_A(\mp\varphi) +\rm{Re}\,D_2 \textbf{e}_B(\mp\varphi) \right.\nonumber\\
&\pm&\left.\rm{Im}\,D_3 \textbf{e}_A(2\varphi_1\pm\varphi) +\rm{Re}\,D_3 \textbf{e}_B(2\varphi_1\pm\varphi)\right],
\end{eqnarray}
\begin{eqnarray}\label{CD123}
  C_1&=&z_{11}^{\ast}z_{23}+z_{12}^{\ast}z_{24}+z_{13}^{\ast}z_{21}+z_{14}^{\ast}z_{22},\nonumber\\
  C_2&=&z_{12}^{\ast}z_{23}+z_{14}^{\ast}z_{21}, C_3=z_{22}^{\ast}z_{13}+z_{24}^{\ast}z_{11},\nonumber\\
  D_1&=&z_{11}^{\ast}z_{21}+z_{12}^{\ast}z_{22}+z_{13}^{\ast}z_{23}+z_{14}^{\ast}z_{24},\nonumber\\
  D_2&=&z_{12}^{\ast}z_{21}+z_{14}^{\ast}z_{23}, D_3=z_{22}^{\ast}z_{11}+z_{24}^{\ast}z_{13}.
\end{eqnarray}
The vector $\textbf{e}_A(\varphi_1)$ is given by Eq.~(\ref{eA}) and
\begin{equation}\label{eB}
    \textbf{e}_B(\varphi_1)=\textbf{e}_{1}\times\textbf{e}_A(\varphi_1)=\textbf{e}_2\sin\varphi_1+\textbf{e}_3\cos\varphi_1.
\end{equation}
Since $\Delta\xi/\Omega\ll 1$, we obtain the mean values
\begin{eqnarray}\label{unimeanvpm}
  \langle\textbf{v}_{\pm}\rangle&=&{\mathcal I}_{\Delta X}(\textbf{v}_{\pm})\nonumber\\
   &=&\pm\textbf{e}_1 \left[J_{-}(|q_1|)\cos 2\alpha + J_{+}(|q_1|)\sin 2\alpha \right]\nonumber\\
   &-&R_v \sin 2\alpha \textbf{e}_B(\mp\varphi),
\end{eqnarray}
\begin{eqnarray}\label{unimeanspm}
  \langle\textbf{s}_{\pm}\rangle&=&{\mathcal I}_{\Delta X}(\textbf{s}_{\pm})\nonumber\\
   &=&\mp\textbf{e}_1 \Sigma_{10}\cos 2\alpha - R_s \sin 2\alpha \textbf{e}_B(\mp\varphi),
\end{eqnarray}
where $\varphi=\delta+2\pi\nu_{pr}t$, $\delta$ specifies the initial precession phase, and $\nu_{pr}=\Delta\xi m_e c^2/h$ is the precession frequency. The transverse components of the precessing vectors of the velocity (probability current density) and the spin are specified by $\alpha$ and the coefficients
\begin{equation}\label{uniRvRs}
    R_v=-\int_0^1 C_{2} dX_4,\quad R_s=-\int_0^1 D_{2} dX_4,
\end{equation}
which depend on $q_1$, as shown in Figs.~\ref{fig12Rv} and \ref{fig13Rs}. The inversion of the quasimomentum $\textbf{q}\mapsto -\textbf{q}$ is described by the replacements $\textbf{v}_{\pm}\mapsto\textbf{v}_{\mp}$ and $\textbf{s}_{\pm}\mapsto\textbf{s}_{\mp}$. It inverts the signs of longitudinal components and reverses the precession directions.

\subsection{\label{sec:bidir}Bidirectional states}
 The bidirectional wave functions $\Psi_1$ and $\Psi_2$~(\ref{unibi}) satisfy the normalization condition $\Psi_{j}^\dag\Psi_{j}=1$. The Hermitian forms $\Psi_j^\dag p_1 \Psi_j$ and $\Psi_j^\dag H \Psi_j$ are both periodic in $X_4$ with the unit period and periodic in $X_1$ with the periods
\begin{equation}\label{bidX1j}
    \Delta X_{1j}=\left|1-(-1)^j q_m\right|^{-1}, q_m=2|q_1|/\Omega, j=1, 2.
\end{equation}
The normalized momentums for the bidirectional states depend on $\alpha$ as follows:
\begin{equation}\label{biPj}
    P_{1j}=\frac{1}{m_{e}c}{\mathcal I}_{\Delta X_{1j}}(\Psi_j^\dag p_1 \Psi_j)= \left[|q_1|-(-1)^j p_{10}\right]\cos 2\alpha,
\end{equation}
where $p_{10}$ is given by Eq.~(\ref{p10}) and
\begin{equation}\label{biIdX}
    {\mathcal I}_{\Delta X_{1j}}(f)\equiv \frac{1}{\Delta X_{1j}}\int_0^1 dX_4 \int_0^{\Delta X_{1j}} f dX_1, j=1, 2.
\end{equation}
The normalized energies
\begin{equation}\label{biEj}
    \frac{1}{m_ec^2}{\mathcal I}_{\Delta X_{1j}}(\Psi_{j}^\dag H \Psi_{j})=E_j
\end{equation}
are independent of $\alpha$ and $\delta$; they are given by Eq.~(\ref{p10}).

The Hermitian forms for the operators of velocity and spin are defined by the relations
\begin{eqnarray}\label{bijS}
  \textbf{j}_j=c\textbf{v}_j=c\sum_{k=1}^3 \textbf{e}_k (\Psi_j^\dag \alpha_k\Psi_j),\nonumber\\
  \textbf{S}_j=\frac{\hbar}{2}\textbf{s}_j=\frac{\hbar}{2}\sum_{k=1}^3 \textbf{e}_k (\Psi_j^\dag \Sigma_k\Psi_j),
\end{eqnarray}
where $j=1, 2$, and
\begin{eqnarray}\label{bivj}
\textbf{v}_1&=&R_1 \textbf{e}_A(\varphi_1)-R_{1214} \textbf{g}_{0-}+R_{1113} \textbf{g}_{2+},\nonumber\\
\textbf{v}_2&=&-R_2 \textbf{e}_A(\varphi_1)+R_{2123} \textbf{g}_{0+}-R_{2224} \textbf{g}_{2-},\nonumber\\
\textbf{s}_1&=&-\frac{1}{2}(1+\Sigma_{10})\textbf{g}_{0-}-I_1 \textbf{g}_{1+}+\frac{1}{2}(1-\Sigma_{10}) \textbf{g}_{2+},\nonumber\\
\textbf{s}_2&=&\frac{1}{2}(1+\Sigma_{10})\textbf{g}_{0+}-I_2 \textbf{g}_{1-}-\frac{1}{2}(1-\Sigma_{10})\textbf{g}_{2-},\nonumber\\
\textbf{g}_{0\pm}&=&\cos2\alpha \textbf{e}_1 \mp\sin2\alpha \textbf{e}_B[\pm(q_m\varphi_1+\delta)],\nonumber\\
\textbf{g}_{1\pm}&=&\sin2\alpha \cos[(1\pm q_m)\varphi_1\pm\delta] \textbf{e}_1 \mp\cos2\alpha \textbf{e}_B(\varphi_1),\nonumber\\
\textbf{g}_{2\pm}&=&\cos2\alpha \textbf{e}_1 \pm\sin2\alpha \textbf{e}_B[(2\pm q_m)\varphi_1\pm\delta].
\end{eqnarray}
At given $q_m, \alpha$, and $\delta$, the scalar coefficients $R_j, I_j, R_{j1j3}$, and $R_{j2j4}$, where $j=1,2$, are periodic in $X_4$ with the unit period. The vectors $\textbf{e}_A, \textbf{g}_{k\pm}$ are independent of $X_4$, but they all have different dependencies on $X_1$. Therefore, the vector functions $\textbf{v}_j=\textbf{v}_j(X_1,X_4)$ and $\textbf{s}_j=\textbf{s}_j(X_1,X_4)$ are periodic in $X_4$ but, in the general case, they are not periodic in $X_1$. However, they become periodic in $X_1$ at some specific values of $q_m$. In particular, the period is equal to $\Delta X_1=2^n$ for $q_m=2^{-n}, n=1, 2,...$, and $\Delta X_1=1$ for any integer $q_m$.

\begin{figure}
\includegraphics{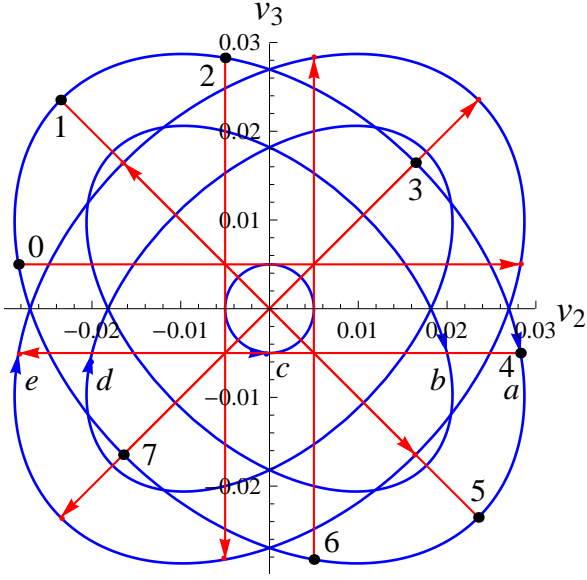}
\caption{\label{fig14}Parametric plot of coordinate curves for $\textbf{v}=v_2\textbf{e}_2+v_3\textbf{e}_3=\textbf{v}_1(X_1,X_4)$ at $q_m=1, \alpha=\pi/4 $, and $\delta=0$: $X_1$ curves $a, b, c, d, e$ for $X_4=0, 1/8, 1/4, 3/8, 1/2$, respectively, $X_1\in [0,1]$; $X_4$ curves with $X_4\in [0,1/2]$ begin at points numbered $k=0,1,...,7$, where $X_1=k/8$.}
\end{figure}

\begin{figure}
\includegraphics{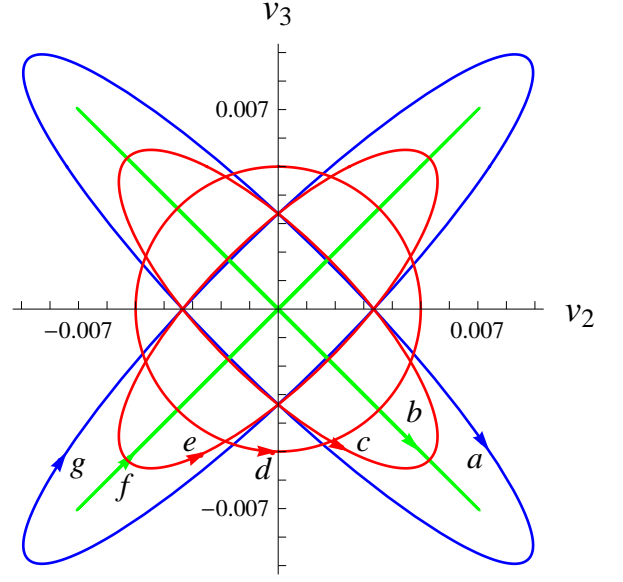}
\caption{\label{fig15}Parametric plot of $X_1$ coordinate curves for  $\textbf{v}=\textbf{v}_1(X_1,X_4)$ in the neighborhood of the instant of time $X_4=1/4$  at $p=1$; $X_4=1/4 + k \delta_a/2, k=-3,-2,-1,0,1,2,3$ for curves $a, b, c, d, e, f, g$, respectively, $\delta_a=0.028125$.}
\end{figure}

\begin{figure}
\includegraphics{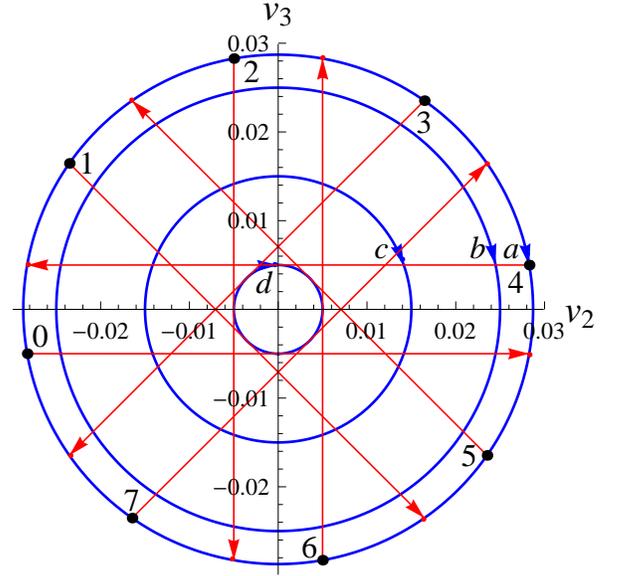}
\caption{\label{fig16} Parametric plot of coordinate curves for  $\textbf{v}=\textbf{v}_2(X_1,X_4)$ at $q_m=1, \alpha=\pi/4 $, and $\delta=0$: $X_1$ curves $a, b, c, d$ for $X_4=0, 1/12, 1/6, 1/4$, respectively, $X_1\in [0,1]$; $X_4$ curves with $X_4\in [0,1/2]$ begin at point numbered $k=0,1,...,7$, where $X_1=k/8$.}
\end{figure}

The relations (\ref{bivj}) define the parametric surfaces $\textbf{v}=\textbf{v}_j(X_1,X_4)$ and $\textbf{s}=\textbf{s}_j(X_1,X_4)$ which can be treated as specific graphic markers of the bispinor wave functions $\Psi_1$ and $\Psi_2$ at given $q_m, \alpha$, and $\delta$. By way of example, let us consider a particular case with $q_m=1, \alpha=\pi/4 $, and $\delta=0$, when $P_{1j}=0$ and $\textbf{e}_1 \cdot \textbf{v}_j\equiv 0, j=1, 2$. In this case, the mean values of momentum with respect to both $\Psi_1$ and $\Psi_2$ are vanishing and the probability streamlines are in the phase planes $X_1=const$. The families of coordinate curves illustrating the dependence of velocity fields $\textbf{v}_1$ and $\textbf{v}_2$ on the spatial coordinate $X_1$ and the time $X_4$ diverge considerably; see Figs.~\ref{fig14}--\ref{fig16}. Unlike $\textbf{v}_1(X_1,X_4)$, the parametric surface $\textbf{v}_2(X_1,X_4)$ has the hole in its center, namely, $|\textbf{v}_2|\geq 0.005$ at all values of $X_1$ and $X_4$; see Fig.~\ref{fig16}. All $X_1$ curves in Fig.~\ref{fig16} are similar in appearance and $X_1$ increases in the clockwise direction, whereas $X_1$ curves in Figs.~\ref{fig14} and \ref{fig15} modify the form with time and reverse their direction at $X_4=1/4$.

\begin{figure}
\includegraphics{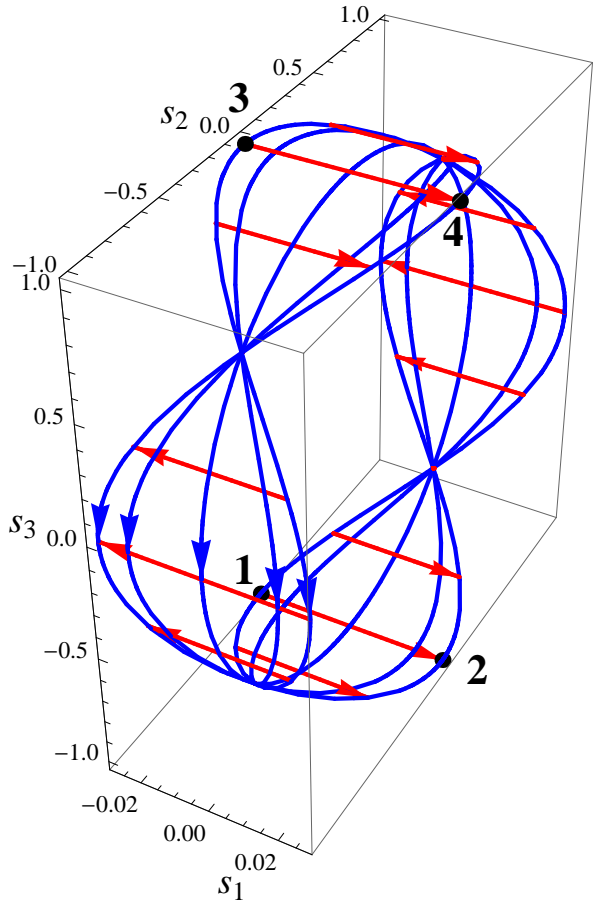}
\caption{\label{fig17} Parametric plot of coordinate curves for  $\textbf{s}=s_1\textbf{e}_1+s_2\textbf{e}_2+s_3\textbf{e}_3=\textbf{s}_1(X_1,X_4)$ at $q_m=1, \alpha=\pi/4 $, and $\delta=0$: $X_1$ curves with $X_1\in [0,1]$ for $X_4=0, -1/4, -1/8, 0, 1/8, 1/4$; $X_4$ curves with $X_4\in [-1/4,1/4]$ for $X_1= k/16, k=0,1,...,15$; $X_1=0$ and $1/2$ for points 1, 2 and 3, 4, respectively;  $X_4=-1/4$ and $1/4$ for points 1, 3 and 2, 4, respectively.}
\end{figure}

\begin{figure}
\includegraphics{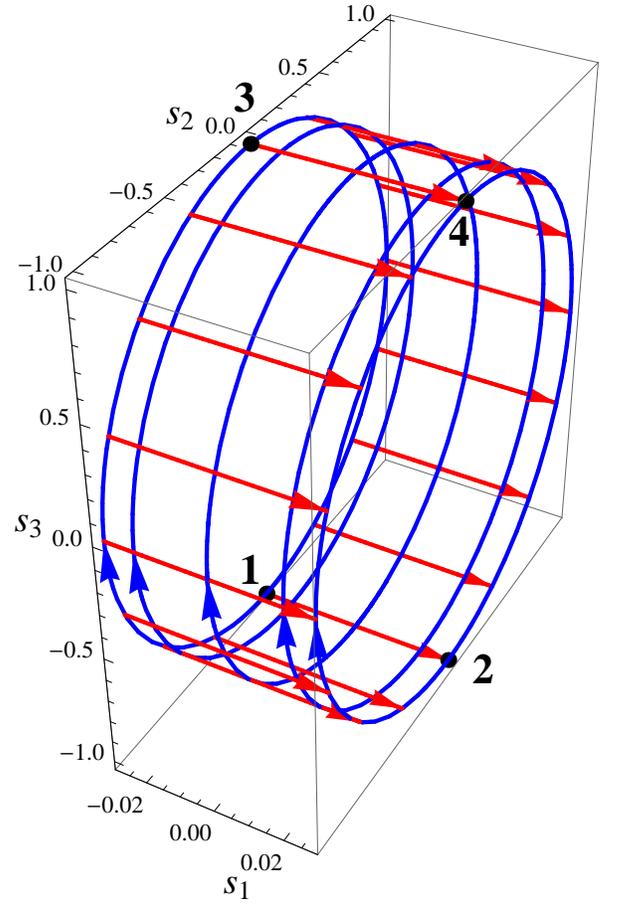}
\caption{\label{fig18} Parametric plot of coordinate curves for  $\textbf{s}=\textbf{s}_2(X_1,X_4)$ at $q_m=1, \alpha=\pi/4 $, and $\delta=0$. The values of $X_1$ and $X_4$ are the same as in Fig.~\ref{fig17}.}
\end{figure}

The Hermitian forms $\textbf{s}_1$ and $\textbf{s}_2$ for the spin operator also diverge considerably; see Figs.~\ref{fig17} and \ref{fig18}. At $q_m=1$ and $\alpha=\pi/4 $, they are described by the relations
\begin{eqnarray}\label{s1d}
  \textbf{s}_1&=&-\frac{1}{2}(1+\Sigma_{10})\textbf{e}_B(-\varphi_1-\delta)-I_1\cos(2\varphi_1+\delta)\textbf{e}_1 \nonumber\\
   &+&\frac{1}{2}(1-\Sigma_{10})\textbf{e}_B(3\varphi_1+\delta),
\end{eqnarray}
\begin{eqnarray}\label{s2d}
  \textbf{s}_2&=&-\frac{1}{2}(1+\Sigma_{10})\textbf{e}_B(\varphi_1+\delta)-I_2\cos\delta \textbf{e}_1 \nonumber\\
   &+&\frac{1}{2}(1-\Sigma_{10})\textbf{e}_B(\varphi_1-\delta).
\end{eqnarray}
The longitudinal component of $\textbf{s}_1$ oscillates with time. The oscillation amplitude depends on $\varphi_1$ and vanishes at points where $\cos(2\varphi_1+\delta)=0$. For the vector $\textbf{s}_2$, the similar oscillation amplitude is independent of $\varphi_1$. It is specified by  $\delta$ and vanishes at $\delta=\pm\pi/2$, but takes maximum value at $\delta=0$, as shown in Fig.~\ref{fig18}.

\section{Conclusion}
The properties of ESTCs vary widely; one can set polarizations, intensities, initial phases of the constitutive electromagnetic plane waves, and the frequency. To construct a specific ESTC, it is useful to evaluate first the structural parameters presented in the Appendix because they specify the
interconnections in the infinite system of matrix equations and, in the final analysis, prescribe the Fourier spectrum of the electron wave function. The presented solutions provide an example of such approach.

At given quasimomenta $\textbf{q}=\textbf{q}_{\pm}\equiv\pm |q_1|\textbf{e}_1$, the Dirac equation in the chiral 2D-ESTC has the four solutions $\Psi_j(\textbf{q}_{\pm}), j=1, 2$, which describe two different spin states of the electron moving along the $X_1$ axis in the positive and negative directions. The bispinor functions $\Psi_j(\textbf{q}_{\pm})=\Psi_j(\textbf{q}_{\pm})(X_1,X_4)$ are uniquely defined by eight complex scalar functions (structural functions) $z_{jk}=z_{jk}(X_4), j=1,2, k=1,2,3,4$, which serve as convenient building blocks of the relations describing the electron properties. These functions are obtained in the form of Fourier expansions, where coefficients can be calculated by making use of the recurrent relations (\ref{Zplus})--(\ref{Zeq0}) and the starting coefficients presented in Figs.~\ref{fig3x210x230}--\ref{fig6dy}.

At any quasimomentum $q_1\neq 0$, the dispersion equation has two solutions which specify wave functions describing electron states with different energy and mean values of momentum and spin operators. The energy level splitting is illustrated in graphical form over a wide range of $q_1$. It is shown that at $|q_1|< q_{10}$, the mean values of velocity and momentum operators are opposite in sign for both of the spin states.

At $q_1\neq 0$ the wave functions $\Psi_j(\textbf{q}_{\pm}), j=1, 2$, form a basis for a  four-dimensional subspace of partial solutions to the Dirac equation, but at $q_1=0$, as a consequence of Eq.~(\ref{limpsi}), this subspace degenerates to the two-dimensional one. In this paper, two families of partial solutions which describe unidirectional and bidirectional states of the Dirac electron are treated. In the comparative analysis of such electron states, it is advantageous to calculate both mean values and Hermitian forms of various operators with respect to the corresponding wave functions, in particular the velocity operator and the spin operator.

The unidirectional electron states are specified by superpositions of two basic wave functions $\Psi_1(\textbf{q}_{\pm})$ and $\Psi_2(\textbf{q}_{\pm})$ corresponding to the same quasimomentum $\textbf{q}_{\pm}$ but describing two different spin states. It is shown that such superpositions describe the electron precession. The magnitudes of transverse components of precessing velocity vectors $\textbf{v}_{\pm}$ and spin $\textbf{s}_{\pm}$ are given by coefficients $R_v$ and $R_s$ depending on $q_1$,  as shown in Figs.~\ref{fig12Rv} and \ref{fig13Rs}.

The bidirectional electron states are specified by superpositions of two basic wave functions $\Psi_j(\textbf{q}_{+})$ and $\Psi_j(\textbf{q}_{-})$ corresponding to the two equal-in-magnitude but oppositely directed quasimomenta and also describing two different spin states. In particular, such superpositions describe the relativistic electron states with the zero mean value of the momentum operator and specific probability current densities and Hermitian forms of the spin operator.

In this paper we present families of nonlocalized solutions of the Dirac equation. They can be used as basis wave functions to construct various localized states of the Dirac electron by applying the general approach proposed in \cite{pre00,*pre01,*pre02}, where it was illustrated for the examples of electromagnetic and weak gravitational fields. Natural crystals prescribe the polarization state and the refractive index of light plane waves and thus provide a means to control the properties of light beams. Similarly, electromagnetic space-time crystals prescribe the spin state and the energy of the Dirac electron. This makes them promising tools to control the quantum states of electrons.

\begin{acknowledgments}
We thank the anonymous referee for important and useful comments, which were used to revise this paper.
\end{acknowledgments}

\appendix*
\section{}
 The definitions of $N_1(m,s)$ and $N_2(s)$ are given in Sec.~\ref{sec:baseq}. Here, we present these major structural parameters in the explicit form that is necessary in any numerical implementation of the general techniques developed in Refs.~\cite{ESTCp1,ESTCp2,ESTCp3,ESTCp4}.

\subsection{Dirac sets of matrices $N_1(m,s)$}
We present $N_1(m,s)$ and $N_2(s)$ in order of the sequential numbering $i=0,1,\dots$ of points $s=s_h(i)\in \mathcal{L}$ (see appendix in Ref.~\cite{ESTCp2}). There are 12 points with $g_{4d}(s)=1$. They are elements (from 2 to 13) of the list
\begin{eqnarray}\label{S69}
 S_{69}=&&\{s_h(i),i=0,1,...,69\}\nonumber\\
 =&&\left\{(0, 0, 0, 0), \right. \nonumber\\
 &&(0, 0, -1, -1),(0, -1, 0, -1),(-1, 0, 0, -1),\nonumber\\
 &&(1, 0, 0, -1),(0, 1, 0, -1),(0, 0, 1, -1),\nonumber\\
 &&(0, 0, -1, 1),(0, -1, 0, 1),(-1, 0, 0, 1),\nonumber\\
 &&(1, 0, 0, 1),(0, 1, 0, 1),(0, 0, 1, 1),\nonumber\\
 &&(0, 0, 0, -2),(0, 0, 0, 2),\nonumber\\
 &&(0, 0, -2, 0),(0, -1, -1, 0),(-1, 0, -1, 0),\nonumber\\
 &&(1, 0, -1, 0),(0, 1, -1, 0),(0, -2, 0, 0),\nonumber\\
 &&(-1, -1, 0, 0),(1, -1, 0, 0),(-2, 0, 0, 0),\nonumber\\
 &&(2, 0, 0, 0),(-1, 1, 0, 0),(1, 1, 0, 0),\nonumber\\
 &&(0, 2, 0, 0),(0, -1, 1, 0),(-1, 0, 1, 0),\nonumber\\
 &&(1, 0, 1, 0),(0, 1, 1, 0),(0, 0, 2, 0),\nonumber\\
 &&(0, 0, -2, -2),(0, -1, -1, -2),(-1, 0, -1, -2),\nonumber\\
 &&(1, 0, -1, -2),(0, 1, -1, -2),(0, -2, 0, -2),\nonumber\\
 &&(-1, -1, 0, -2),(1, -1, 0, -2),(-2, 0, 0, -2),\nonumber\\
 &&(2, 0, 0, -2),(-1, 1, 0, -2),(1, 1, 0, -2),\nonumber\\
 &&(0, 2, 0, -2),(0, -1, 1, -2),(-1, 0, 1, -2),\nonumber\\
 &&(1, 0, 1, -2),(0, 1, 1, -2),(0, 0, 2, -2),\nonumber\\
 &&(0, 0, -2, 2),(0, -1, -1, 2),(-1, 0, -1, 2),\nonumber\\
 &&(1, 0, -1, 2),(0, 1, -1, 2),(0, -2, 0, 2),\nonumber\\
 &&(-1, -1, 0, 2),(1, -1, 0, 2),(-2, 0, 0, 2),\nonumber\\
 &&(2, 0, 0, 2),(-1, 1, 0, 2),(1, 1, 0, 2),\nonumber\\
 &&(0, 2, 0, 2),(0, -1, 1, 2), (-1, 0, 1, 2),\nonumber\\
 &&\left. (1, 0, 1, 2),(0, 1, 1, 2), (0, 0, 2, 2)\right\}.
\end{eqnarray}

The $D$ sets of matrices $N_1[m,s_h(i)], i=1,\dots,12$, have the form~\cite{ESTCp3}
\begin{eqnarray*}
  D_s&&\left\{N_1[m,(0,0,-1,-1)]\right\}=\\
     &&\left\{-2(A_{31}w_1+A_{32}w_2), 0,i A_{32}\Omega, -i A_{31}\Omega,0,0,0,0,\right.\\
     &&\left. 0,0, -A_{31}\Omega_{-},-A_{32}\Omega_{-},0,0,0,0\right\},
\end{eqnarray*}
\begin{eqnarray*}
  D_s&&\left\{N_1[m,(0,-1,0,-1)]\right\}=\\
     &&\left\{-2(A_{21}w_1+A_{23}w_3),i A_{21}\Omega,-i A_{23}\Omega,0,0,0,0,0,\right.\\
     &&\left. 0, -A_{23}\Omega_{-},-A_{21}\Omega_{-},0,0,0,0,0\right\},
\end{eqnarray*}
\begin{eqnarray*}
  D_s&&\left\{N_1[m,(-1,0,0,-1)]\right\}=\\
     &&\left\{-2(A_{12}w_2+A_{13}w_3),-i A_{12}\Omega,0,i A_{13}\Omega,0,0,0,0,\right.\\
     &&\left. 0, -A_{13}\Omega_{-},0,-A_{12}\Omega_{-},0,0,0,0\right\},
\end{eqnarray*}
\begin{eqnarray*}
  D_s&&\left\{N_1[m,(1,0,0,-1)]\right\}=\\
     &&\left\{-2(A_{42}w_2+A_{43}w_3),i A_{42}\Omega,0,-i A_{43}\Omega,0,0,0,0,\right.\\
     &&\left. 0, -A_{43}\Omega_{-},0,-A_{42}\Omega_{-},0,0,0,0\right\},
\end{eqnarray*}
\begin{eqnarray*}
  D_s&&\left\{N_1[m,(0,1,0,-1)]\right\}=\\
     &&\left\{-2(A_{51}w_1+A_{53}w_3),-i A_{51}\Omega,i A_{53}\Omega,0,0,0,0,0,\right.\\
     &&\left. 0, -A_{53}\Omega_{-},-A_{51}\Omega_{-},0,0,0,0,0\right\},
\end{eqnarray*}
\begin{eqnarray*}
  D_s&&\left\{N_1[m,(0,0,1,-1)]\right\}=\\
     &&\left\{-2(A_{61}w_1+A_{62}w_2),0,-i A_{62}\Omega,i A_{61}\Omega,0,0,0,0,\right.\\
     &&\left. 0,0,-A_{61}\Omega_{-},-A_{62}\Omega_{-},0,0,0,0\right\},
\end{eqnarray*}
\begin{eqnarray*}
  D_s&&\left\{N_1[m,(0,0,-1,1)]\right\}=\\
     &&\left\{-2(A^{\ast}_{61}w_1+A^{\ast}_{62}w_2),0,i A^{\ast}_{62}\Omega,-i A^{\ast}_{61}\Omega,0,0,0,0,\right.\\
     &&\left. 0,0, -A^{\ast}_{61} \Omega_{+},-A^{\ast}_{62}\Omega_{+},0,0,0,0\right\},
\end{eqnarray*}
\begin{eqnarray*}
  D_s&&\left\{N_1[m,(0,-1,0,1)]\right\}=\\
     &&\left\{-2(A^{\ast}_{51}w_1+A^{\ast}_{53}w_3),i A^{\ast}_{51}\Omega,-i A^{\ast}_{53}\Omega,0,0,0,0,0,\right.\\
     &&\left. 0, -A^{\ast}_{53}\Omega_{+}, -A^{\ast}_{51}\Omega_{+},0,0,0,0,0\right\},
\end{eqnarray*}
\begin{eqnarray*}
  D_s&&\left\{N_1[m,(-1,0,0,1)]\right\}=\\
     &&\left\{-2(A^{\ast}_{42}w_2+A^{\ast}_{43}w_3),-i A^{\ast}_{42}\Omega,0,i A^{\ast}_{43}\Omega,0,0,0,0,\right.\\
     &&\left. 0,-A^{\ast}_{43}\Omega_{+},0, -A^{\ast}_{42}\Omega_{+},0,0,0,0\right\},
\end{eqnarray*}
\begin{eqnarray*}
  D_s&&\left\{N_1[m,(1,0,0,1)]\right\}=\\
     &&\left\{-2(A^{\ast}_{12}w_2+A^{\ast}_{13}w_3),i A^{\ast}_{12}\Omega,0,-i A^{\ast}_{13}\Omega,0,0,0,0,\right.\\
     &&\left. 0,-A^{\ast}_{13}\Omega_{+},0, -A^{\ast}_{12}\Omega_{+},0,0,0,0\right\},
\end{eqnarray*}
\begin{eqnarray*}
  D_s&&\left\{N_1[m,(0,1,0,1)]\right\}=\\
     &&\left\{-2(A^{\ast}_{21}w_1+A^{\ast}_{23}w_3),-i A^{\ast}_{21}\Omega,i A^{\ast}_{23}\Omega,0,0,0,0,0,\right.\\
     &&\left. 0,-A^{\ast}_{23}\Omega_{+}, -A^{\ast}_{21}\Omega_{+},0,0,0,0,0\right\},
\end{eqnarray*}
\begin{eqnarray*}
  D_s&&\left\{N_1[m,(0,0,1,1)]\right\}=\\
     &&\left\{-2(A^{\ast}_{31}w_1+A^{\ast}_{32}w_2),0,-i A^{\ast}_{32}\Omega,i A^{\ast}_{31}\Omega,0,0,0,0,\right.\\
     &&\left. 0,0,-A^{\ast}_{31}\Omega_{+}, -A^{\ast}_{32}\Omega_{+},0,0,0,0\right\}.
\end{eqnarray*}

\subsection{Coefficients $N_2(s)$ }
There are 56 points $s=s_h(i)\in \mathcal{L}, i=13,\dots,68$ with $g_{4d}(s)=2$. They are elements (from 14 to 69) of the list $S_{69}$. The list of the coefficients $N_2(s)$ has the form
\begin{eqnarray*}
   \{N_2[s_h(i)],i=13,\dots,68\} =\\
   \left\{2 \left(A_{12} A_{42} + A_{13} A_{43} + A_{21} A_{51} + A_{23} A_{53}\right.\right.\\
   \left. + A_{31} A_{61} + A_{32} A_{62}), \right.\\
   2 \left(A^{\ast}_{12} A^{\ast}_{42} + A^{\ast}_{13} A^{\ast}_{43} + A^{\ast}_{21} A^{\ast}_{51} + A^{\ast}_{23} A^{\ast}_{53}\right.\\
   \left. + A^{\ast}_{31} A^{\ast}_{61} + A^{\ast}_{32} A^{\ast}_{62}\right),\\
   2 (A_{31} A^{\ast}_{61} + A_{32} A^{\ast}_{62}), 2 (A_{31} A^{\ast}_{51} + A_{21} A^{\ast}_{61}),\\
   2 (A_{32} A^{\ast}_{42} + A_{12} A^{\ast}_{62}), 2 (A^{\ast}_{12} A_{32} + A_{42} A^{\ast}_{62}),\\
   2 (A^{\ast}_{21} A_{31} + A_{51} A^{\ast}_{61}), 2 (A_{21} A^{\ast}_{51} + A_{23} A^{\ast}_{53}),\\
   2 (A_{23} A^{\ast}_{43} + A_{13} A^{\ast}_{53}), 2 (A^{\ast}_{13} A_{23} + A_{43} A^{\ast}_{53}),\\
   2 (A_{12} A^{\ast}_{42} + A_{13} A^{\ast}_{43}), 2 (A^{\ast}_{12} A_{42} + A^{\ast}_{13} A_{43}),\\
   2 (A_{13} A^{\ast}_{23} + A^{\ast}_{43} A_{53}), 2 (A^{\ast}_{23} A_{43} + A^{\ast}_{13} A_{53}),\\
   2 (A^{\ast}_{21} A_{51} + A^{\ast}_{23} A_{53}), 2 (A_{21} A^{\ast}_{31} + A^{\ast}_{51} A_{61}),\\
   2 (A_{12} A^{\ast}_{32} + A^{\ast}_{42} A_{62}), 2 (A^{\ast}_{32} A_{42} + A^{\ast}_{12} A_{62}),\\
   2 (A^{\ast}_{31} A_{51} + A^{\ast}_{21} A_{61}), 2 (A^{\ast}_{31} A_{61} + A^{\ast}_{32} A_{62}),\\
   (A_{31} + i A_{32}) (A_{31} - i A_{32}), 2 A_{21} A_{31}, 2 A_{12} A_{32},\\
   2 A_{32} A_{42}, 2 A_{31} A_{51}, (A_{21} + i A_{23}) (A_{21} - i A_{23}),\\
   2 A_{13} A_{23}, 2 A_{23} A_{43}, (A_{12} + i A_{13}) (A_{12} - i A_{13}),\\
   (A_{42} + i A_{43}) (A_{42} - i A_{43}), 2 A_{13} A_{53}, 2 A_{43} A_{53},\\
   (A_{51} + i A_{53}) (A_{51} - i A_{53}), 2 A_{21} A_{61}, 2 A_{12 }A_{62},\\
   2 A_{42} A_{62}, 2 A_{51} A_{61}, (A_{61} + i A_{62}) (A_{61} - i A_{62}),\\
   (A^{\ast}_{61} + i A^{\ast}_{62}) (A^{\ast}_{61} - i A^{\ast}_{62}), 2 A^{\ast}_{51} A^{\ast}_{61}, 2 A^{\ast}_{42} A^{\ast}_{62},\\
   2A^{\ast}_{12} A^{\ast}_{62}, 2 A^{\ast}_{21} A^{\ast}_{61}, (A^{\ast}_{51} + i A^{\ast}_{53}) (A^{\ast}_{51} - i A^{\ast}_{53}),\\
   2 A^{\ast}_{43} A^{\ast}_{53}, 2 A^{\ast}_{13} A^{\ast}_{53}, (A^{\ast}_{42} + i A^{\ast}_{43}) (A^{\ast}_{42} - i A^{\ast}_{43}),\\
   (A^{\ast}_{12} + i A^{\ast}_{13}) (A^{\ast}_{12} - i A^{\ast}_{13}), 2 A^{\ast}_{23} A^{\ast}_{43}, 2 A^{\ast}_{13} A^{\ast}_{23},\\
   (A^{\ast}_{21} + i A^{\ast}_{23}) (A^{\ast}_{21} - i A^{\ast}_{23}), 2 A^{\ast}_{31} A^{\ast}_{51}, 2 A^{\ast}_{32} A^{\ast}_{42},\\
   \left.  2 A^{\ast}_{12} A^{\ast}_{32}, 2 A^{\ast}_{21} A^{\ast}_{31}, (A^{\ast}_{31} + i A^{\ast}_{32}) (A^{\ast}_{31} - i A^{\ast}_{32})\right\}.
\end{eqnarray*}

\bibliography{BorzdovGN_2016_a}

\end{document}